\documentclass[Afour,sageh,times]{sagej}

\usepackage{moreverb}

\newcommand\BibTeX{{\rmfamily B\kern-.05em \textsc{i\kern-.025em b}\kern-.08em
T\kern-.1667em\lower.7ex\hbox{E}\kern-.125emX}}

\def\volumeyear{2019}

\usepackage{url}
\usepackage{cleveref}
\crefname{lstlisting}{listing}{listings}
\Crefname{lstlisting}{Listing}{Listings}
\usepackage{mathtools}

\usepackage[ruled,norelsize]{algorithm2e}
\makeatletter
\newcommand{\removelatexerror}{\let\@latex@error\@gobble}
\makeatother

\usepackage{listings}
\usepackage{graphicx}
\usepackage{subfig}
\usepackage{caption}
\usepackage{wrapfig}
\usepackage{xcolor}
    \definecolor{ListingsKeywordColor}{rgb}{0,0,0.4}
    \definecolor{ListingsIdentifierColor}{rgb}{0,0.5,0}
    \definecolor{ListingsCommentColor}{rgb}{0.4,0.4,0.4}
    \definecolor{ListingsStringColor}{rgb}{0.6000,0.3333,0.7333}
    \definecolor{ListingsRuleSepColor}{rgb}{0,0,0}
    \definecolor{ListingsEmphColor}{rgb}{0,0.6667,0.6667}
    \definecolor{ListingsBreakSymbolColor}{rgb}{0.780,0.082,0.522}
    \definecolor{LinkColor}{rgb}{0,0,0.5}
    \definecolor{UnitColor}{rgb}{0,0,0}
    \definecolor{MathsVectorColor}{rgb}{0,0,0}
    \definecolor{MathsMatrixColor}{rgb}{0,0,0}
    \definecolor{MyGreen}{HTML}{228B22}
    \definecolor{MyBlue}{HTML}{0000FF}
    \colorlet{MatrixElementsLight}{gray!40!white}
    \colorlet{MatrixElementsDark}{gray!80}
    \colorlet{MyGreenLight}{MyGreen!40!white}
    \colorlet{MyGreenDark}{MyGreen!80}    
    \colorlet{MyBlueLight}{MyBlue!40!white}
    \colorlet{MyBlueDark}{MyBlue!80}    
    \colorlet{MyRedLight}{red!20!white}
    \colorlet{MyRedDark}{red!60}

\def\twosum{\texttt{twosum}}
\def\twoprod{\texttt{twoprod}}
\def\binary{\texttt{binary64}}
\newcommand{\fma}{{\tt fma}\xspace}
\def\binary{\texttt{binary64}}
\def\blas{BLAS}
\def\exblas{ExBLAS}
\def\exsum{{\sc exsum}}
\def\axpy{{\sc axpy}}

\def\dotp{{\sc dot}}
\def\exdotp{{\sc exdot}}

\def\CC{{ C\nolinebreak[4]\hspace{-.05em}\raisebox{.4ex}{\scriptsize\bf ++ }}}

\newcommand{\becomes}{:=}

\usepackage{amsmath}

\renewcommand{\dotp}{{\sc dot}\xspace}
\renewcommand{\axpy}{{\sc axpy}\xspace}
\newcommand{\axpyl}{{\sc axpy}(-type)\xspace}

\newcommand{\spmv}{{\sc SpMV}\xspace}

\newcommand{\bd}{e}

\newcommand{\tA}{\tilde{A}}
\newcommand{\td}{\tilde{d}}
\newcommand{\mytw}{\tilde{w}}
\newcommand{\trho}{\tilde{\rho}}

\newcommand{\R}{\mathbb{R}}

\newcommand{\proc}[1]{#1}
\newcommand{\modeA}[1]{#1}
\newcommand{\step}[1]{{\it #1}}

\begin{document}

\title{Reproducibility of Parallel Preconditioned Conjugate Gradient in Hybrid Programming Environments}

\author{Roman Iakymchuk\affilnum{1,2},
Maria Barreda\affilnum{3},
Stef Graillat\affilnum{1},  
Jos\'e I. Aliaga\affilnum{3},
Enrique S. Quintana-Ort\'i\affilnum{4}
}

\affiliation{\affilnum{1}Sorbonne Universit\'e, LIP6, France\\
\affilnum{2}Fraunhofer ITWM, Germany\\ 
\affilnum{3}Universitat Jaime I, Spain\\
\affilnum{4}Universitat Polit\`ecnica de Val\`encia, Spain
}

\corrauth{Roman Iakymchuk, Sorbonne Universit\'e, LIP6, Campus Pierre et Marie Curie,
4 place Jussieu,
75252 PARIS CEDEX 05, France}
\email{roman.iakymchuk@sorbonne-universite.fr}

\begin{abstract}
The Preconditioned Conjugate Gradient method is often employed for the solution of linear systems of equations arising in numerical simulations of physical phenomena. While being widely used, the solver is also known for its lack of accuracy while computing the residual. In this article, we propose two algorithmic solutions that originate from the ExBLAS project to enhance the accuracy of the solver as well as to ensure its reproducibility in a hybrid MPI + OpenMP tasks programming environment. One is based on ExBLAS and preserves every bit of information until the final rounding, while the other relies upon floating-point expansions and, hence, expands the intermediate precision. Instead of converting the entire solver into its ExBLAS-related implementation, we identify those parts that violate reproducibility/non-associativity, secure them, and combine this with the sequential executions. These algorithmic strategies are reinforced with programmability suggestions to assure deterministic executions. Finally, we verify these approaches on two modern HPC systems: both versions deliver reproducible number of iterations, residuals, direct errors, and vector-solutions for the overhead of less than 37.7\,\% on 768 cores.
\end{abstract}

\keywords{
Preconditioned Conjugate Gradient, MPI, OpenMP tasks, reproducibility, accuracy, floating-point expansion, long accumulator, fused multiply-add.
}

\maketitle

\section{Introduction}
%
%
Many current scientific and engineering problems involve 
the solution of
large and sparse linear systems of equations. 
Some traditional examples appear, for example, in circuit and device simulation, 
quantum physics, large-scale eigenvalue computations, 
nonlinear sparse equations, and all sorts of applications 
that include the discretization of 
partial differential equations (PDEs)~\cite{barrettemplates}.
For many problems (especially those associated with 3-D models), 
the size and complexity of these systems have turned iterative 
projection methods, based on Krylov subspaces, into 
a highly competitive approach compared with direct solvers~\cite{Saa03}.
In particular, the Conjugate Gradient (CG) method is 
one of the most efficient Krylov subspace-based algorithms 
for the solution of sparse linear systems when the coefficient matrix 
is symmetric positive definite (s.p.d.)~\cite{Saa03}.
Preconditioning is usually incorporated in real implementations 
of the method in order to accelerate the convergence of the method and improve its numerical
features, yielding the Preconditioned Conjugate Gradient (PCG)  method.

One would expect that the results of different runs of 
PCG are identical, for instance, in the number of iterations, the intermediate and final residuals, as well as the solution-vector. However, in practice this is not often 
the case due to different reduction trees -- the Message Passing Interface (MPI)
implementations (libraries)~\cite{gropp2014using} offer up to 14 different implementations for reduction --, 
OpenMP tasks scheduling, data alignment, instructions used, etc. Each of these factors may change the execution order of floating-point operations, which are commutative but non-associative, and, hence, result in non-reproducible results. We define {\em reproducibility as the ability to obtain a bit-wise identical and  accurate result for multiple executions on the same data}. 
Therefore, our aim of this study is to ensure reliable computations (we also refer to them as robust), at reasonable cost, for codes that 
leverage PCG (or any similar Krylov subspace solver) and encounter numerical issues during sensitive computations of the residual. Our approach and routines are also aimed to be used for {\em debugging} as we ensure reproducible residuals, direct errors, number of iterations, and solution-vector from the sequential, pure MPI, and even hybrid MPI + OpenMP versions.

Ensuring the bit-wise reproducibility is often a complex and expensive task that imposes modifications to the algorithm and its underlying parts such as the BLAS (Basic Linear Algebra Subprograms) routines~\cite{Lawson79BLAS1,Dongarra90BLAS3}. These modifications are necessary to preserve every bit of information (both result and error)~\cite{Collange15Parco} or, alternatively, to cut off some parts of the data and operate on the remaining most-significant parts~\cite{ozblas,Demmel14OneRed}. Furthermore, the bit-wise reproducibility can become expensive with the overhead of at least 8\,\% for parallel reduction~\cite{Collange15Parco,Demmel14OneRed}, up to 2x-4x for matrix-vector product~\cite{Iakymchuk19ReproLU}, and more than 10x for matrix-matrix multiplication~\cite{Iakymchuk16Gemm}.
%
%
In this paper, we aim to revisit reproducibility and raise its appeal through reducing its negative impact on performance and minimizing changes to both the algorithm and its building blocks. We also raise a question: {\em Can reproducibility of algorithms be ensured by design with both minimal changes to algorithms and almost negligible overhead?} 
Hence, our idea is to address those parts of algorithms that violate associativity -- such as parallel reductions, dot products, and possible replacements by compilers of $a*b+c$ in favor of fused multiply-add (\fma) operation, etc. -- as well as to combine that with sequential executions of sub-blocks/subroutines. Such sequential execution of operations is reproducible under some constraints, for example the same initial conditions on the input data like data alignment.

We consider to verify this idea (both algorithmic and programmability) on a typical sparse linear 
algebra solver such as PCG and ensure its reproducibility on parallel distributed-memory systems using a hybrid combination of
the MPI + OpenMP-tasks programming models. On one hand, the hybridization reduces the communication burden being more focused on inner node computations and work balancing, especially on nodes with large core counts such as those in the MareNostrum4 platform at
\textit{Barcelona Supercomputing Center}. On the other hand, it introduces a new challenge in the form of a double-level reduction: an initial reduction among tasks inside a process/node, followed by one among processes. Thus, we ensure reproducibility of the PCG solver by preventing non-deterministic executions as follows:
\begin{itemize}
  \item We construct two reproducible solutions: a first one on the ExBLAS approach~\cite{exblas} and an alternative lightweight version based on floating-point expansions (FPEs). The ExBLAS-based approach with its cornerstone Kulisch long accumulator~\cite{K08} is robust but expensive since it is designed to cover severe (ill-conditioned) cases with very broad dynamic ranges. Motivated by ``100 bits suffice for many HPC applications" as noted by David Bailey at ARITH-21~\cite{baileyarith21} and  a mini accumulator from the ARM team~\cite{hpanchor17,hpanchor19}, we derive a faster but less generic version using FPEs, which is the other core algorithmic component in the ExBLAS approach, aiming to adjust the algorithm to the problem at hand.
  \item As a consequence, we also address the common issue of sparse iterative solvers 
-- the accuracy while computing the residual -- 
and propose to use solutions that offer reproducibility 
(and potentially correct-rounding) only while computing the corresponding dot products.
  \item Hence, we derive two hybrid (MPI + OpenMP tasks), reproducible, and accurate dot products using ExBLAS and FPEs.
  \item Finally, we demonstrate applicability and feasibility of the aforementioned idea with the ExBLAS- and FPE-based approaches in the hybrid 
  MPI + OpenMP implementation of PCG on an example of a 3D Poisson's equation with 27 stencil points as well as several test matrices from the SuiteSparse matrix collection. This extends our previous results with the pure MPI implementation of PGC~\cite{iakymchuk19jcam} to the more complex double-level dot products and reductions with dynamic scheduling of the tasks. 
\end{itemize}
{\em To sum up}: the FPE-based (we also call it Opt) solution is efficient and fast, but it is limited to cases where the condition number and/or the dynamic range do not exceed certain thresholds, e.g. the dynamic range is below $10^{50}$. 
(At this point, we note that the condition of a linear system can be cheaply estimated with fair accuracy.)
In comparison, the ExBLAS-based solution is reserved for extreme cases as well as problems where we do not have any information about the problem at hand.

This article is organized as follows. 
\Cref{sec:background} reviews several aspects of computer arithmetic, 
in particular floating-point expansion and 
long accumulator, as well as the \exblas\ approach for accurate and reproducible computations. \Cref{sec:algs} introduces the PCG algorithms and describes in details its hybrid (MPI+OpenMP) implementation. We present strategies for ensuring reproducibility of PCG in~\Cref{sec:repro} and evaluate corresponding implementations in~\Cref{sec:results}. Finally, \Cref{sec:related:works} reviews related work, while \Cref{sec:conclusion} draws conclusions and outlines future directions.

\section{Background}
\label{sec:background}
At first, we briefly introduce the floating-point arithmetic that
consists in approximating real numbers by numbers that have a finite, fixed-precision representation. These numbers are composed of a significand, an exponent, and a sign:
$x = \pm \underbrace{x_0 . x_1 \ldots x_{M-1}}_{mantissa} \times b^{e}, \,\, 0 \leq x_i \leq b-1, \,\, x_0 \neq 0,$
where $b$ is the  basis ($2$ in our case), $M$ is the precision, and $e$ stands for the exponent that is bounded ($e_{\min} \leq e \leq e_{\max}$).

The IEEE 754 standard~\cite{IEEE7542008}, created in 1985 and then revised in 2008, has
led to a considerable enhancement in the reliability of numerical computations by
rigorously specifying the properties of floating-point arithmetic. This standard
is now adopted by most processors, thus leading to a much better
portability of numerical applications.
The standard specifies floating-point formats, which are often associated with precisions like {\em binary16}, {\em binary32}, and {\em binary64}, see~\Cref{tb:ieee754}.
Floating-point representation allows numbers to 
cover a wide 
\textit{dynamic range} that
is defined as the absolute ratio between the number with the largest 
magnitude and the number with the smallest non-zero magnitude in a set. For instance, \binary\ (double-precision) can represent 
positive numbers from $4.9\times10^{-324}$ to $1.8\times10^{308}$, so it covers a dynamic range of $3.7\times10^{631}$.
\begin{table*}
\centering
\caption{Parameters for three IEEE arithmetic precisions; quadruple (128 bits) is omitted.}
\label{tb:ieee754}
\begin{tabular}{llllll}
\hline
Type & Size & Significand & Exponent & Rounding unit & Range \\
\hline\noalign{\vskip .5mm} 
half & 16 bits & 11 bits & 5 bits & $u = 2^{-11} \approx 4.88 \times 10^{-4}$ & $\approx 10^{\pm 5}$ \\
single & 32 bits & 24 bits & 8 bits & $u = 2^{-24} \approx 5.96 \times 10^{-8}$ & $\approx 10^{\pm 38}$\\
double & 64 bits & 53 bits & 11 bits & $u = 2^{-53} \approx 1.11 \times 10^{-16}$ & $\approx 10^{\pm 
308}$\\
\hline
\end{tabular}
\end{table*}

The IEEE 754 standard requires correctly rounded results for the basic arithmetic operations $(+, -, \times , /, \sqrt{~},$ {\tt fma}$)$. It 
means that the operations are performed as if the result was first computed with an infinite precision and then rounded 
to the floating-point format. The correct rounding criterion guarantees a unique, well-defined answer, ensuring bit-wise reproducibility for a single operation.
Several rounding modes are provided. The standard also contains the reproducibility clause that forwards the reproducibility issue to language standards.  
Emerging attention to reproducibility strives to draw more careful attention to the problem by the computer arithmetic community. It has led to the inclusion of error-free 
transformations (EFTs) for addition and multiplication -- to return the exact outcome as the result and the error -- to assure numerical reproducibility of floating-point operations,     
into the revised version of the standard. These mechanisms, once implemented in hardware, will simplify our reproducible algorithms -- like the ones used in the ExBLAS~\cite{Collange15Parco}, ReproBLAS~\cite{Demmel14OneRed}, OzBLAS~\cite{ozblas} libraries -- and boost their performance.

There are three approaches that enable the addition of floating-point numbers without incurring round-off errors or with reducing their impact. 
The main idea is to keep track of both the result and the errors during the course of computations.
The first approach uses EFT to compute both the result and the rounding error and stores them in a floating-point expansion (FPE), 
which is an unevaluated sum of $p$ floating-point numbers, 
whose components are ordered in magnitude with minimal overlap to cover the whole range of exponents. 
Typically, FPE relies upon the use of the traditional EFT for addition that is \twosum~\cite{Knu69} (\Cref{alg:TwoSum}) and for multiplication that is \twoprod\ EFT~\cite{Ogita05accuratesum} (\Cref{alg:TwoProd}). Note that the underlying architecture should support {\tt fma}, which is often the case. Otherwise, we refer to Algorithm 3.3 in~\cite{Ogita05accuratesum}, which relies on Dekker's algorithm for splitting a floating-point number~\cite{Dek71}; this altogether requires 17 flops in contrary to 3 flops of \Cref{alg:TwoProd} with {\tt fma}. 
%
\begin{minipage}[t]{0.495\textwidth}
\vspace{-3.2pt}
 \removelatexerror
 \begin{algorithm}[H]
   \caption{Error-free transformation for the summation of two floating-point numbers.}
   \label[algorithm]{alg:TwoSum}
    \SetKwProg{Fn}{Function}{}{}
    \KwIn{$a,b$ are two floating-point numbers.}
    \KwOut{$r,s$ are the result and the error, resp.}
    \Fn{$[r, s]$ = \texttt{twosum}\,($a, b$)} {
      $r \becomes a + b$\\
      $z \becomes r - a$\\
      $s \becomes (a - (r - z)) + (b - z)$
    }
 \end{algorithm}
\end{minipage}
\begin{minipage}[t]{0.495\textwidth}
\vspace{-3.2pt}
 \removelatexerror
 \begin{algorithm}[H]
   \caption{Error-free transformation for the product of two floating-point numbers.}
   \label[algorithm]{alg:TwoProd}
    \SetKwProg{Fn}{Function}{}{}
    \KwIn{$a,b$ are two floating-point numbers.}
    \KwOut{$r,s$ are the result and the error, resp.}    
    \Fn{$[r, s]$ = \texttt{twoprod}\,($a, b$)} {
      $r \becomes a * b$\\
      $s \becomes fma(a, b, -r)$\\
    }
 \end{algorithm}
\end{minipage}

The second approach projects the finite range of exponents of floating-point numbers into a long vector so called 
a long (fixed-point) accumulator and stores every bit there. For instance, Kulisch~\cite{Kulisch11} proposed to use a 4288-bit long accumulator for the exact dot product of two vectors composed of \binary\ numbers; such a large long accumulator is designed to cover all the severe cases without overflows in its highest digit.

The third approach is based on slicing or splitting a floating-point number into slices using Dekker's algorithm. Then, the same work is carried separately on each slice and the accumulated results are aggregated/merged. More details and analysis can be found in~\cite{RuOgOi2008a}, with ideas originating from~\cite{ZielkeDrygalla03}. This approach is implemented in ReproBLAS and OzBLAS.

\subsection{ExBLAS -- Exact BLAS}
\label{sec:exblas}
The ExBLAS project~\cite{Iakymchuk15ExBLAS} is an effort to derive fast, accurate, and reproducible BLAS library
by constructing  a multi-level approach for these operations that are tailored for various modern architectures with their
complex multi-level memory structures.
On one side, this approach aims to ensure similar performance compared with the non-deterministic parallel
counterparts. On the other side, the approach preserves every bit of information before the final rounding to 
the desired format to assure correct-rounding and, therefore, reproducibility.
Hence, ExBLAS combines together long accumulator and FPE into algorithmic solutions. In addition, it efficiently tunes and implements them on various architectures, including conventional CPUs, NVIDIA and AMD GPUs, and Intel Xeon Phi co-processors (for details we refer to~\cite{Collange15Parco}).
Thus, \exblas\ assures reproducibility through assuring correct-rounding.

The cornerstone of ExBLAS is the reproducible parallel reduction, which is at the core of many BLAS routines.
The ExBLAS parallel reduction relies upon FPEs with the \twosum\ EFT~\cite{Knu69} and long accumulators, so it is correctly rounded and reproducible. In practice, the latter is invoked only once per overall summation which results in the little overhead (less than $8$\,\%) on accumulating large vectors. Our interest in this article is the dot product of two vectors, which is another crucial fundamental \blas\ operation. The \exdotp\ algorithm is based on the previous \exsum\ algorithm and 
the \twoprod\ EFT~\cite{Ogita05accuratesum} (see~\Cref{alg:TwoProd}): the algorithm accumulates the result and the error of \twoprod\ to same FPEs and then follows the \exsum\ scheme. These and the other routines -- such as matrix-vector product, triangular solve, and matrix-matrix multiplication -- are distributed in the ExBLAS library\footnote{ExBLAS repository: \url{https://github.com/riakymch/exblas}.}. In this paper, we derive a hybrid MPI + OpenMP tasks \exdotp, where a long accumulator is shared among OpenMP tasks within one process and each OpenMP thread owns two FPEs underneath (one for the result and the other for the error) that are merged at the end of computations .

\section{Algorithm(s)}
\label{sec:algs}
In this section we review the PCG algorithm and its task-parallel implementation using MPI and OpenMP tasks.
The goal of the following analysis is twofold: to offer a complete description of the parallelization approach and, even more 
important, to identify key inter-node (i.e., between MPI ranks) and intra-node (i.e., between threads executing tasks) 
communications, in  particular reductions, which pose a challenge to ensuring reproducibility.

\subsection{Preconditioned Conjugate Gradient Solver}
We consider the linear system $Ax=b$, where 
the coefficient matrix $A \in \mathbb{R}^{n \times n}$ is sparse and symmetric positive definite (s.p.d.),
with $n_z$ nonzero entries;
$b \in \mathbb{R}^n$ is the right-hand side vector; and $x \in \mathbb{R}^n$ is the sought-after solution vector.
Figure~\ref{fig:pcg} presents the algorithmic description of the classical iterative PCG. In the body loop of the algorithm, the following operations are executed: a sparse matrix-vector product (\spmv) (S1), three DOT products (S2,S6, and S8), three AXPY (-like) operations (S3, S4, and S7), the preconditioner application (S5), and a few scalar operations~\cite{barrettemplates}.
\begin{figure*}
\centering
\hspace{-24pt}\begin{minipage}{0.5\textwidth}
{\small
\begin{tabular}{|l|}
\hline
   Compute preconditioner for $A \rightarrow M$ 
\\ Set starting guess $x^{(0)}$ 
\\ Initialize $z^{(0)}, d^{(0)}, \beta^{(0)}, \tau^{(0)}, l:= 0$ (iteration count)     
\\ \hline
\vspace{-2ex}
\\ $r^{(0)} :=b-Ax^{(0)}$                         
\\ $\tau^{0}:= <r^{(0)}, r^{(0)}>$ 
\\ {\bf while} $(\tau^{(l)} > \tau_{\max})$              
\\
\begin{minipage}{.95\textwidth}
\[
\begin{array}{l|l@{~}c@{~}l|l}
  \multicolumn{1}{l}{\mbox{\rm Step}}  &  \multicolumn{3}{l}{\mbox{\rm Operation}} & \mbox{\rm Kernel} 
\\\hline
  \step{S1}: &   w^{(l)}&:=&Ad^{(l)}                         & \mbox{\rm \spmv}
\\\step{S2}: &   \rho^{(l)}&:=&\beta^{(l)}/{<d^{(l)},w^{(l)}>}       & \mbox{\rm \dotp~product}
\\\step{S3}: &   x^{(l+1)}&:=&x^{(l)}+\rho^{(l)} d^{(l)}           & \mbox{\rm \axpy}
\\\step{S4}: &   r^{(l+1)}&:=&r^{(l)}-\rho^{(l)} w^{(l)}           & \mbox{\rm \axpy}
\\\step{S5}: &   z^{(l+1)}&:=&M^{-1} r^{(l+1)}           & \mbox{\rm Apply precond.}
\\\step{S6}: &   \beta^{(l+1)}&:=& <z^{(l+1)}, r^{(l+1)}>  & \mbox{\rm \dotp ~product}
\\
  \step{S7}: &   d^{(l+1)}&:=& (\beta^{(l+1)}/\beta^{(l)}) d^{(l)} + z^{(l+1)}& \mbox{\rm \axpy-like}
\\\step{S8}: &   \tau^{(l+1)}&:=& <r^{(l+1)}, r^{(l+1)}>   & \mbox{\rm \dotp ~product}
\\           &   l&:=&l+1                               
\end{array}
\]
\end{minipage}
\\ {\bf end while}                                     
\\ \hline
\end{tabular}
}
\vspace{-4pt}
\caption{Formulation of the PCG solver annotated with computational kernels. The threshold
$\tau_{\max}$ is an upper bound on the relative residual for the computed approximation to the solution.
In the notation, 
$<\cdot,\cdot>$ computes the \dotp\ (inner) product of its vector arguments.}
\vspace{4pt}
\label{fig:pcg}
\end{minipage}
\hspace{5mm}
\begin{minipage}{0.42\textwidth}
\vspace{-26pt}
{\small
\begin{tabular}{|l|}
\hline
   Compute preconditioner for $\modeA{A} \rightarrow M$ 
\\ Set starting guess $x$ 
\\ Initialize $z, d, \beta, \tau, l:= 0$     
\\ \hline
\vspace{-2ex}
\\ $\modeA{r} :=\modeA{b}-\modeA{A}x$                        
\\ $\tau:= <r, \modeA{r}>$ 
\\ {\bf while} $(\tau > \tau_{\max})$ 
\\
\begin{minipage}{0.9\textwidth}
\[
\begin{array}{l|l@{~}c@{~}l|l}
  \multicolumn{1}{l}{\mbox{\rm Step}}  &  \multicolumn{3}{l}{\mbox{\rm Operation}} & \mbox{\rm Communication}
\\\hline
             &   \beta'&:=& \beta                          & \mbox{\rm --}
\\ \hline 
  \step{S1}: &  &                                                                & 
\\  ~~\step{S1.1}: &   d &\rightarrow& \bd                                       & \mbox{\rm Allgatherv}
\\  ~~\step{S1.2}: &   \modeA{w}&:=&\modeA{A}\bd                                 & \mbox{\rm --}
\\\hline
  \step{S2}: &   \rho&:=&\beta/{<d,\modeA{w}>}                                   & \mbox{\rm Allreduce}
\\\hline
  \step{S3}: &   x&:=&x+\rho d                                                   & \mbox{\rm --}
\\\step{S4}: &   \modeA{r}&:=&\modeA{r}-\rho \modeA{w}                                   & \mbox{\rm --}
\\\step{S5}: &   z&:=&M^{-1} \modeA{r}                                           & \mbox{\rm --}
\\ \hline
  \step{S6}: &   \beta&:=& <z,\modeA{r}>                                         & {\mbox{\rm Allreduce}}
\\\step{S8}: &   \tau&:=& <r, \modeA{r}>                                         & {\mbox{\rm Allreduce}}
\\\hline           
   \step{S7}: &   d&:=& (\beta/\beta') d + z                                     & \mbox{\rm --}
\\           &   l&:=&l+1                                                        & 
\end{array}
\]
\end{minipage}
\\ {\bf end while}                                    
\\ \hline
\end{tabular}
}
\vspace{-4pt}
\caption{Message-passing formulation of the PCG solver annotated with communication.}
\vspace{-10pt}
\label{fig:pcg-comm}
\end{minipage}
\end{figure*}
In particular, in the proposed implementation of the PCG method, we incorporate a Jacobi preconditioner~\cite{Saa03}, which is composed of the elements in the diagonal of the matrix ($M := D = diag(A)$). Therefore, 
the application of the preconditioner is carried out on a vector and involves an element-wise multiplication of two vectors.

\subsection{Message-passing PCG}
\label{sec:mpipcg}
In this subsection we analyse the communication patterns of a  message-passing implementation of the PCG solver that operates in a distributed-memory platform.
For clarity, hereafter we will drop the superindices that denote the iteration count in the variable names. 
The following considerations are taken into account in the analysis of the communications:
 \begin{itemize}
   \vspace{-3.pt}
   \setlength\itemsep{-1.5pt}
\item The parallel platform comprises
      $K$ processes (or MPI ranks), denoted as $\proc{P}_1$, $\proc{P}_2$, \ldots, $\proc{P}_{K}$. 
\item The coefficient matrix $A$ is partitioned into $K$ blocks of rows ($A_1$, $A_2$, $\hdots$, $A_k$), 
with the $k$-th {\em distribution block} $A_k \in \R^{p_k \times n}$ 
stored in $\proc{P}_k$, and $n=\sum_{k=1}^{K}p_k$.
\item Vectors are partitioned and allocated 
conformally with the block-row distribution of $A$. For example, the residual vector $r$ is partitioned as $r_1$, $r_2$, $\hdots$, $r_K$,
where $\proc{P}_k$ stores $r_k$.
\item The scalars $\alpha, \beta, \rho, \tau$ are replicated on all $K$ processes.
 \vspace{-3.pt}
\end{itemize}

Considering these previous aspects, we next examine how they affect the different computational kernels (\step{S1}--\step{S8}) that are executed in a single PCG iteration in Figure~\ref{fig:pcg}.

{\it Sparse matrix-vector product (\step{S1}):}
The input operands are the coefficient matrix $A$, which is distributed by blocks of rows, and the vector $d$, which is partitioned and distributed according to $A$.
A communication stage is required before executing this kernel in order to assemble the distributed parts of vector $d$ into a single vector $\bd$, which is replicated in all processes. We denote this communication as $d \rightarrow \bd$, which can be performed in MPI via an \texttt{MPI\_Allgatherv}. Note that vector $\bd$ is the only array that is replicated in all processes. After that, the computation can proceed in parallel and each process calculates its local slice of the output vector $w$: $\proc{P}_k: w_k \becomes A_k \, \bd.$

{\it\dotp\ products (\step{S2}, \step{S6}, \step{S8}):}
In this kernel, each process can compute concurrently a partial result (in step \step{S2}, $\proc{P}_k$ calculates $\rho_k \becomes <d_k,w_k>$).
Then, these intermediate values are reduced into a globally-replicated scalar (for example,
$\rho := \beta/(\rho_1+\rho_2+\cdots+\rho_{K})$ in \step{S2}). We implement this reduction in MPI using \texttt{MPI\_Allreduce}. Applying this idea to all the \dotp~products, there are three process synchronizations because $\rho, \beta, \tau$ are globally-replicated.

{\it \axpyl vector updates (\step{S3}, \step{S4}, \step{S7}):}
The \axpy~kernel involves two distributed vectors ($x$ and $d$ in \step{S3}) and a globally-replicated scalar ($\rho$ in \step{S3}).
This kernel can be executed concurrently because all processes can perform their local parts of the computation without any communication ($\proc{P}_k: x_k \becomes x_k + \rho \, d_k$).

{\it Application of the preconditioner (\step{S5}):}
The kernel in step \step{S5} consists in applying the Jacobi preconditioner $M$. In order to do that, vector $r$ is scaled by the diagonal of the matrix. Here, each process stores a different group of the diagonal elements and also a local piece of the vector $r$, so that, the computations can be done in parallel, i.e $\proc{P}_k: z_k \becomes M^{-1}_k r_k$.

The algorithm with communications is summarized in~\Cref{fig:pcg-comm}. We can re-arrange the operations to reduce the the number of synchronizations in the loop body of the PCG solver, as shown there. Concretely, pushing up step \step{S8} next to step \step{S6}, we can simultaneously execute these two reductions by merging them into one reduction and, hence, the number of synchronizations decreases from three to two per iteration of PCG. 

\subsection{Task-parallelism in message-passing PCG}
\label{subsec:tp-pcg}
In a cluster of multicore processors, a good practice to increase the performance of the codes is to introduce an additional level of parallelism. This level is exploited in each node of the cluster using, for example, OpenMP. The analysis in~\cite{AliBFBQ17,Barreda19} exposes that, in the PCG, a reasonable option is to leverage {\em task-parallelism}, which consists in dividing each kernel into a collection of finer-grain operations, or tasks. Then, each thread executes a different task and two consecutive kernels can be executed concurrently avoiding a thread-synchronization point after each kernel, as described next.

In the following analysis, for simplicity, we merge
the execution of \step{S3} with that of \step{S4};
and \step{S8} with \step{S6}. Therefore, we will only consider kernels
\step{S1}--\step{S2} and \step{S4}--\step{S7} 
in the loop body of the PCG solver (see Figure~\ref{fig:pcg-comm}).
Thus, the operations in the solver are interlaced by a series of data dependencies which
impose a strict order of execution:

\vspace*{-2ex}
\begin{footnotesize}
\[
      \underbracket{\cdots \step{S7}}_{\textrm{iteration}~l-1} \xrightarrow{d} 
      \underbracket{\step{S1} \xrightarrow{w}
      \step{S2} \xrightarrow{\rho} 
      \step{S4} \xrightarrow{r}
      \step{S5} \xrightarrow{z} 
      \step{S6} \xrightarrow{\beta} 
      \step{S7}}_{\textrm{iteration}~l} \xrightarrow{d} 
      \underbracket{\step{S1} \cdots}_{\textrm{iteration}~l+1},
\]
\end{footnotesize}
\noindent
where the variable that generates the dependency is denoted on top of each dependency arrow.

Exploiting task-parallelism allows that some of these kernels can be (partially) computed concurrently (denoted
with the symbol ``$\|$''), breaking the strict inter-kernel barriers due to the dependencies;
in particular, we aim to attain a parallel execution with
\step{S1} $\|$ \step{S2} and \step{S4} $\|$ \step{S5} $\|$ \step{S6}. 


\paragraph{Sparse matrix-vector product \step{S1} $\|$ \dotp product \step{S2}:}
On the one hand, the local operands to process $\proc{P}_k$ of the \spmv can be divided as $w_k := A_k$ \, \bd : 

\vspace*{1ex}
$\left[
\renewcommand{\arraystretch}{1.2}
\begin{array}{c}
          \mytw_1
\\ \hline \mytw_2
\\ \hline \vdots
\\ \hline \mytw_I
\end{array}\right] :=
\left[
\renewcommand{\arraystretch}{1.2}
\begin{array}{c|c|c|c}
          \tA_{1,1} & \tA_{1,2} & \ldots & \tA_{1,J}
\\ \hline \tA_{2,1} & \tA_{2,2} & \ldots & \tA_{2,J}
\\ \hline \vdots    & \vdots    & \ddots & \vdots
\\ \hline \tA_{I,1} & \tA_{I,2} & \ldots & \tA_{I,J}
\end{array}\right] 
\left[
\renewcommand{\arraystretch}{1.2}
\begin{array}{c}%
\bd_1
\\ \hline \bd_2
\\ \hline \ddots
\\ \hline \bd_J
\end{array}\right].$
\vspace*{2ex}

\noindent
Here, we can consider each group of rows as a task, which computes the corresponding \spmv operation to obtain a partial result, $\mytw_k$. 
For example, if we consider $\mytw_1$,
there is a task calculating $\mytw_1 = \sum_{j=1}^J \tA_{1,j} e_j$.

On the other hand, the computation local to
$\proc{P}_k$ for the \dotp product \step{S2}
can be decomposed into $S$ tasks. These tasks can be computed
concurrently by partitioning the input operands $d_k,w_k$ into $S$ pieces, with each task obtaining a partial result $\trho_s$:
\[
\rho_k := ~<d_k,w_k>
\equiv 
\trho_s := ~<\td_s,\mytw_s>,  s=1,2,\ldots,S.
\]
These partial results are reduced to generate a unique value local to the $k$-th 
node, $\rho_k := \sum_{s=1}^S \trho_s$, and these local values are thereafter reduced across all ${K}$ nodes to produce the globally replicated scalar $\rho := \sum_{k=1}^{K} \rho_k$.

Note that the advantage here is that we can eliminate the dependency \step{S1} $\rightarrow$ \step{S2}, by splitting up these operations into fine-grain tasks. Hence, the execution of some tasks of the second kernel can start as soon as the corresponding results of 
the previous one are available, resulting in a partially-parallel execution of these two tasks. 
However, the global reduction required at the end of \step{S2} enforces a task/process synchronization point that is an impediment to extend this idea further that point.

\paragraph{\axpy vector update \step{S4} $\|$ preconditioner application \step{S5} $\|$ \dotp product \step{S6}:}
\step{S4}--\step{S6} can be computed in parallel by applying
a similar division of the three kernels into fine-grain. Nevertheless, again a task/process synchronization is required right after \step{S6}.

\paragraph{\axpy vector update \step{S7} and \spmv~\step{S1} (subsequent iteration):}
The convergence test and the requirement to perform the replication $d \rightarrow \bd$ at the beginning of each iteration, 
inserts a process synchronization that makes impossible
the concurrent computation of the local tasks corresponding to these two kernels.

\subsection{Implementation using MPI+OpenMP}

In this subsection we detail how to exploit the described two levels of parallelism via a combination of two 
parallel programming interfaces: MPI~\cite{mpi} and OpenMP~\cite{openmp}.

We leverage OpenMP tasks to implement task-parallelism. At execution time, the runtime system underlying OpenMP detects data dependencies between tasks, with the help of compiler directives 
(\texttt{\#pragma omp task}) 
annotated with clauses that indicate the task operands' directionality (input ({\tt in}), output ({\tt out}) or both ({\tt inout})). Then, a task graph is generated during the execution, which is used to schedule the tasks to the cores, exploiting the inherent task-level parallelism while fulﬁlling the dependencies embedded in the graph.

As an example, the \dotp product, which computes $\alpha := x^Ty$, $x,y\in\R^q$, is annotated as

\vspace*{1ex}
{\footnotesize
\begin{verbatim}
      #pragma omp task depend(in:x[0:n], y[0:n]) 
                       depend(out:alpha)
      ddot (int q, double *x, int incx, 
           double *y, int incy, double alpha);
\end{verbatim}
}

For the routine \axpy $y := y+\alpha x$, the code snippet using OpenMP tasks is as follows:

\vspace*{1ex}
{\footnotesize
\begin{verbatim}
      #pragma omp task depend(in:alpha, x[0:n]) 
                       depend(inout:y[0:n]) 
      daxpy (int q, double alpha, double *x,
             int incx, double *y, int incy);
\end{verbatim}
}
The replication of  vector $d$ into $e$ is performed across the processes using 
the MPI collective \texttt{MPI\_Allgatherv}, as stated previously. 
To ensure that all the processes have finalized their computation of $d$ prior to the MPI collective, we introduce a task barrier, 
using the directive \texttt{\#pragma omp taskwait}.
This creates a task synchronization point because it enforces that all tasks up to that point are completed. Furthermore, this syncronization point is leveraged to perform the convergence test ($\tau > \tau_{\max}$?) right after it, 
which is followed by an implicit MPI syncronization across processes in the MPI collective primitive.

The \texttt{MPI\_Allreduce} primitive is used to implement the global reductions. Similarly to the previous case, 
we insert a \texttt{\#pragma omp taskwait} on the specific variable being reduced 
before invoking the MPI collectives for the reduction. This ensures 
that all tasks operating on that variable have been finalized prior the reduction across
nodes can start.
Moreover, atomic updates are employed to accumulate the results from each reduction task (e.g., $\tilde{\beta}_n$ for \step{S8}) into the local result ($\beta_k$).

\begin{figure}
\begin{center}
\includegraphics[height=0.6\textheight,width=\columnwidth]{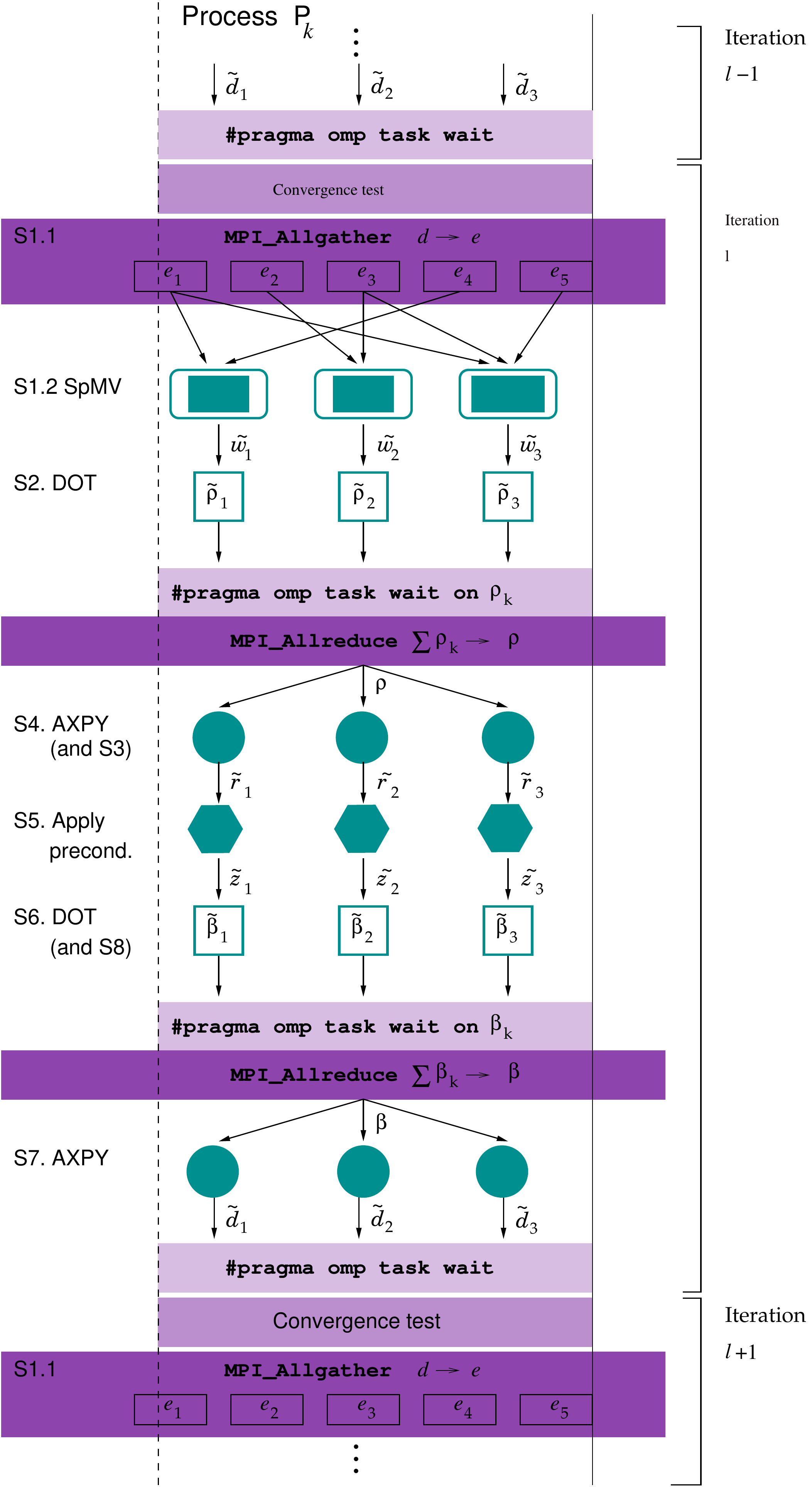}
\caption{Dependencies between kernels in the PCG solver.}
\label{fig:pcg-itergraph}
\end{center}
\end{figure}

The previous description is condensed in Figure~\ref{fig:pcg-itergraph}, focusing on the operations computed by the process
$\proc{P}_k$, during the iteration $l$. 

\section{Reproducibility of PCG}
\label{sec:repro}
In this section, we present our strategies for ensuring reproducibility of the PCG solver.
The first strategy relies on the ExBLAS approach, while the second is derived from it and is based on FPEs. 
Both strategies are reinforced with programmability components such as the explicit use of \fma\ instructions and 
a careful re-arrangement of computations. 
Therefore, the reproducibility of the PCG solver is guaranteed via reproducibility of its building blocks on each iteration. 

\subsection{\exblas-based Strategy}
\label{sec:exblas-based}
\Cref{sec:exblas} provides an overview of the \exblas\ approach. Here we exploit the \exblas\ parallel reduction in conjunction with the \twoprod\ EFT to derive a hybrid MPI (inter-node, distributed) and OpenMP (intra-node) for the \dotp\  products appearing in PCG. The intra-node \dotp\ product is presented here, and its distributed part is described in~\Cref{sec:repropcg} together with the FPE-based alternative. 

For accurate and reproducible \dotp\ product within an MPI process, we rely upon OpenMP tasks following the ExBLAS approach. We allocate one long accumulator per MPI process as well as, within the \texttt{exblas::cpu::exdot}, a vector of FPEs shared among OpenMP threads. Hence, the work on the process-local \dotp\ products is divided into multiple task (more than threads) in such a way that the intermediate results from each task are stored in this large vector of FPEs. To complete the local \dotp\ products we flush all FPEs sequentially into the process-owned long accumulator. \Cref{lst:dot-omp-exblas} outlines the code snippet of this implementation. \texttt{Accumulate} is presented in Algorithm~\ref{alg:expansion}, however here it also includes a possibility to flush the error to the long accumulator in case of not enough capacity to store this error.
\begin{minipage}[t]{0.49\textwidth}
\begin{lstlisting}[language=C,caption=Process-local \dotp\ product with OpenMP tasks and ExBLAS., label={lst:dot-omp-exblas}]
double *fpe = (double *) calloc(NBFPE*omp_get_num_threads(), sizeof(double));
for (int i = 0; i < N_k; i += bm ) {
   int cs = N_k - i;
   int c = cs < bm ? cs : bm;

   #pragma omp task depend(in:a[i:i+c-1], b[i:i+c-1]) depend(out:fpe) firstprivate(i,c) 
   for(int j = i; j < (i+c); j++) {
       double r1;
       double x = TwoProductFMA(a[j],b[j],r1);
       Accumulate(&fpe[omp_get_thread_num()*NBFPE], x);
       Accumulate(&fpe[omp_get_thread_num()*NBFPE], r1);
    }
}
#pragma omp taskwait
for (int i=0; i < omp_get_num_threads(); i++)
    Flush(&fpe[i*NBFPE]);
\end{lstlisting}
\end{minipage}

Delivering both correctly rounded and reproducible results, \exblas\ has two major drawbacks~\cite{Collange15Parco}. The first drawback is related to the required memory storage, which amounts for 
$n_t\times p + acc_s,$
where $n_t$ is the thread count, $p$ is the size of floating-point expansion, and $acc_s$ is the size of superaccumulator (2,098 bits for summation). The second drawback is the number of  required operations: For an input vector of size $n$ with dynamic range  $d$, the cost of accumulation is
\begin{eqnarray*}
 &n \times\left\lceil{\frac{d}{52}}\right\rceil \times \frac{C_{fpe}}{VL} + n_t \times p \times VL \times 
C_{sa},\,\mbox{when}\,\,\, d < 52 p,\\
 &n \times \left( p \times \frac{C_{fpe}}{VL} + C_{sa} \right) + n_t \times p \times VL \times 
C_{sa},\,\mbox{otherwise},
\end{eqnarray*}
where $C_{fpe} = 6$ flops, see~\Cref{alg:TwoSum}, is the cost of the expansion update, $VL$ 
is the architecture-dependent vector length on SIMD architectures (4 with AVX and 1 on GPUs), and $C_{sa} = 16$ 
flops $+2$ indirect memory accesses is the cost of the long accumulator update. The right-hand side term is the cost of 
flushing expansions to long accumulators at the end of the summation and gets negligible as $n$ increases. These two drawbacks can be observed for compute-intensive kernels, leading to large performance overheads~\cite{Iakymchuk16Gemm}. However, these drawbacks are either hardly visible or relatively small on bandwidth- and memory-bound operations such as the \dotp\ product (reduction) and potentially PCG due to the possibility to saturate bandwidth and hide the cost of extra computations and memory needs.

\subsection{FPE-based Strategy}
\label{sec:fpe-based}
We introduce a lightweight strategy for reproducibility using the \exblas\ approach as a starting point. 
The ExBLAS drawbacks served us as a motivation to design an alternative, cheaper strategy for reproducible computations with accuracy (correct rounding) guarantees. 
Examining the PCG method for moderately conditioned but largely sparse matrices, like the studied Poisson matrices, 
we come to the conclusion that the method can successfully accommodate accurate and reproducible computations, ensuring their robustness, using eight-floating point numbers, meaning the FPE of size 8 (FPE8). In fact, the size of FPEs is tunable and exposed to the end-user.  This approach is complemented with the early-exit technique~\cite{Collange15Parco}: we stop propagating zero-errors in the FPE. From our experience, the early-exit technique significantly improves performance. According to~\cite{Hida01}, FPE8 is capable to represent at least $424=8\cdot 53$ bits of significant. 

Our main motivation for iterative solvers, where next iteration corrects the previous estimate, is to provide a good enough associativity-assuring approach since the properties provided by ExBLAS get demolished by the next computation/iteration. In fact, we are working on developing a concept of weak reproducibility~\cite{cre19} -- reproducibility under a certain accuracy guarantee, e.g. defined as the input tolerance, that is not necessarily correct rounding. However, we want all computations on a single iteration to be reproducible.  Therefore, 
the FPEs of size 8 with the early-exit technique is generic enough to cover a wide range of problems with various condition numbers and/or dynamic ranges. We also use two different FPEs underneath (one for results and another for errors) that are merged at the end of computations before rounding.

We discuss here the \dotp\ product using OpenMP tasks, while the distributed \dotp\ product is presented in the section below. 
\Cref{lst:dot-omp-fpe} outlines the FPE-based solution: each MPI process allocates a vector of FPEs for each OpenMP thread and invokes a local routine to conduct \dotp\ products on their local copies of vectors of size $N_k$. This local \dotp\ product routine subdivides the process-local \dotp\ product into tasks of size $bm$; each task calls a sequential \dotp\ product. This \dotp\ product as in~\Cref{lst:dot-omp-exblas} is composed of the call to \twoprod\ EFT (\Cref{alg:TwoProd}) for the exact multiplication of two floating-point numbers; and, then, the accumulation of the output result and the error to the thread local FPEs with the help of Algorithm~\ref{alg:expansion}, which relies upon the \twosum\ EFT (\Cref{alg:TwoSum}). 
Later, the FPEs with the result and the error are combined into one by invoking Algorithm~\ref{alg:fpesum}, which calls Algorithm~\ref{alg:expansion} in a loop over the FPE with errors: 
$\mathrm{Accumulate}(fpe,fperr[i])$. To complete the OpenMP \dotp\ product, we perform the process-local reduction on FPEs by sequentially executing Algorithm~\ref{alg:fpesum}. Finally, we round the FPE-result to the target precision using the {\tt NearSum} algorithm~\cite{RuOgOi2008b}, which is described in~\Cref{sec:repropcg}.
\begin{minipage}[t]{0.48\textwidth}
\begin{lstlisting}[language=C,caption=Process-local \dotp\ product with OpenMP tasks and FPEs., label={lst:dot-omp-fpe}]
void bblas_ddot(int bm, int N_k, double *X, double *Y, double *results)
{
    for (int i=0; i<m; i+=bm ) {
        int cs = m - i;
        int c = cs < bm ? cs : bm;
        #pragma omp task depend(in:X[i:i+c-1], Y[i:i+c-1]) depend(out:results) firstprivate(i,c,m)
        dot(c, X, Y, i, i, &results[NBFPE * omp_get_thread_num()]);
    }
    #pragma omp taskwait
    for (int i=1; i < omp_get_num_threads(); i++)
        fpeSum(&results[0], &results[NBFPE*i], NBFPE);
}
\end{lstlisting}
\end{minipage}

\noindent
\begin{minipage}[t]{.48\textwidth}
 \removelatexerror
 \begin{algorithm}[H]
   \caption{Aggregation of two FPEs of size $p$.}
   \label{alg:fpesum}
    \SetKwProg{Fn}{Function}{}{}
    \Fn{{\tt fpesum}$(a,b,p)$} {
      \KwIn{$b$ is a FPE.}
      \KwOut{$a$ is a FPE containing the result.}
      \For{$i=0 \to p-1$}{
          Accumulate(a,b[i])
      }
    }  
 \end{algorithm}
\end{minipage}

\noindent
\begin{minipage}[t]{0.48\textwidth}
\removelatexerror
 \begin{algorithm}[H]
   \caption{Adding a floating-point number $x$ to a floating-point expansion $a$ of size $p$.}
   \label{alg:expansion}
    \SetKwProg{Fn}{Function}{}{}
    \Fn{$\mathrm{Accumulate}(a,x)$} {
      \KwIn{$x$ is a floating-point number.}
      \KwOut{$a$ is a FPE containing the result.}
      \For{$i=0 \to p-1$}{
          $(a[i], x) \becomes \texttt{twosum}(a[i], x)$
      }
    }  
 \end{algorithm}
\end{minipage}

\subsection{Re-installing Reproducibility of PCG}
\label{sec:repropcg}
We re-assure reproducibility of parallel PCG by first examining potential sources of non-deterministic computations and, 
in addition, presenting our mitigation strategies for them. Note that we target a hybrid MPI + OpenMP tasks implementation of PCG, 
where each process conducts computations on its own local slices of the matrix as well as the vectors (see~\Cref{sec:mpipcg} and \Cref{fig:pcg-comm}).

{\it\dotp\ products (\step{S2}, \step{S6}, \step{S8}):}
The main issue of non-determinism emerges from \dotp\ products and, thus, the parallel reductions such as \texttt{MPI\_Allreduce()} that are employed in order to compute the tolerance $\tau$ as well as both $\beta$ and $\rho$. Hence, we 1) exploit the \exblas\ approach to build reproducible and correctly-rounded \dotp\ product; 2) construct \dotp\ product solely based on FPEs; 3) extend  the \exblas- and FPE-based \dotp\ products to distributed memory in order to make them suitable for the PCG algorithm in~\Cref{fig:pcg-comm}. 
While \Cref{sec:exblas-based,sec:fpe-based} present implementations of \dotp\ product using OpenMP tasks, i.e. within each MPI process, \Cref{lst:allreduceexblas,lst:allreducefpe} provide pseudo-codes for our implementation of the distributed \dotp\ product using the \exblas\ and lightweight strategies, respectively. After carrying out process-local \dotp\ products, via either \exblas- or FPE-based implementations, we realise the global reduction by splitting them into three stages:
\begin{itemize}
	\item \texttt{MPI\_Reduce()} acting on either long accumulators or FPEs. For the ExBLAS approach, since the long accumulator is an array of long integers, we apply regular reduction. Note that we may need to carry an extra intermediate normalization after the reduction of $2^{K-1}$ long accumulators, where $K=64-52=12$ is the number of carry-safe bits per each digit of long accumulator. For the FPE approach, we may need to renormalize FPEs using the Priest's renormalization method~\cite{Hida01,Priest91} and define the MPI operation that is based on the \twosum\ EFT, see Algorithm~\ref{alg:fpesum};
	\item Rounding to double: for long accumulators, we use the \exblas-native \texttt{Round()} routine. To guarantee correctly rounded results of the FPE-based computations, we employ the \texttt{NearSum} algorithm from~\cite{RuOgOi2008b} for FPEs of size eight or variable size; it may require renormalization before. 
	\item \texttt{MPI\_Bcast()} to distribute the result of \dotp\ product to the other processes as only master performs rounding.
\end{itemize}
Splitting the \texttt{MPI\_Allreduce()} operation into \texttt{MPI\_Reduce()} and \texttt{MPI\_Bcast()} provides us full control of the operation and even may lead to better performance as noted in~\cite{reprompi2016}.

\noindent
\begin{minipage}[t]{0.48\textwidth}
\begin{lstlisting}[language=C,caption=Reproducible Allreduce with ExBLAS., label={lst:allreduceexblas}]
std::vector<int64_t> h_superacc(BIN_COUNT);
exblas::exdot (..., &h_superacc[0]);
exblas::Normalize (&h_superacc[0]);
MPI_Reduce ((myId==0)?MPI_IN_PLACE:&h_superacc[0], &h_superacc[0], BIN_COUNT, MPI_LONG, MPI_SUM, ...);
if (myId == 0) {
  beta = exblas::Round (&h_superacc[0]);
}
MPI_Bcast (&beta, 1, MPI_DOUBLE, ...);
\end{lstlisting}
\end{minipage}

\noindent
\begin{minipage}[t]{0.48\textwidth}
\begin{lstlisting}[language=C,caption=Reproducible Allreduce with FPEs only., label={lst:allreducefpe}]
double *fpes = (double *) calloc(NBFPE*omp_get_num_threads(), sizeof(double));
bblas_ddot (..., &fpes[0]);
renormalize(&fpes[0]);    // optional
MPI_Op Op;    // user-defined reduction operation
MPI_Op_create (fpesum, 1, &Op);
MPI_Reduce ((myId==0)?MPI_IN_PLACE:&fpes[0], &fpes[0], N, MPI_DOUBLE, Op, ...);
if (myId == 0) {
  beta = Round (&fpes[0]);   // NearSum
}
MPI_Bcast (&beta, 1, MPI_DOUBLE, ...);
\end{lstlisting}
\end{minipage}


{\it Sparse matrix-vector product (\step{S1}):}
The other reproducibility issue is hidden in the computation of the sparse matrix-vector product. With the current distributed implementation of this operation, each MPI process computes its dedicated part $w_k$ of the vector $w$ by multiplying a block of rows $A_k$ by the vector $e$. These process-local multiplications are correspondingly divided into tasks, where each task is responsible for a product of a sub-block of rows by the vector. Since the computations are carried locally and sequentially, they are deterministic and, thus, reproducible. However, some parts of the code like $a+=b*c$ -- present in the original implementation of PCG -- may not always provide with the same result, depending on the compiler optimization strategies.


Our approach to solve this issue is to explicitly instruct compilers to use \fma\footnote{This and the other case of $y=a*x+b*y$ are analyzed in more details in~\cite{reprofeltor}.}. Note that the underlying architecture should support \fma; otherwise, this may lead to the runtime error. This is possible through the \texttt{std::fma} instruction added to the \CC11 language standard. With this option, we avoid non-determinism in the order of operations, reduce the number of rounding errors from three to two, and, therefore, achieve reproducibility for this type of operations. Consequently, we accomplish reproducibility for the sparse matrix-vector multiplication. 

{\it\axpyl vector updates (\step{S3}, \step{S4}, \step{S7}):}
For this type of operations, we rely upon the sequential MKL implementation of \axpyl. Alternatively, we can replace this call to MKL's \axpyl by our implementation using \fma to ensure correctly-rounded and, hence, reproducible results. This will not impact performance since the algorithm is strictly memory-bound and this type of kernels are not performance-crucial.

{\it Application of the preconditioner (\step{S5}):}
The application of the Jacobi preconditioner is rather simple: first, the inverse of the diagonals are computed and then the application of the preconditioner only involves element-wise multiplication of two vectors. Thus, this part is both correctly rounded and reproducible.

{\it Reproducibility and accuracy of both approaches:}
It is evident that the results provided by \exblas\ \dotp\ are both correctly-rounded and reproducible. With the lightweight \dotp, we search for the minimal size of FPE such that we still preserve every bit of both the result and the error. For the studied 3D Poisson's equation, the sweet spot is the FPE of size $3$, which ensures identical results to \exblas\ and the reference  highly accurate solution. 
%
However, we aim also to be generic and, hence, we provide the implementation that relies on FPEs of size eight with the early-exit technique. 
We add a check for the FPE-based implementation for those cases where the condition number and/or the dynamic range are too large and we cannot keep every bit of information. A warning is then raised, 
offering also a suggestion to switch to the ExBLAS-based implementation. 
Nonetheless, 
note that the lightweight implementation is intended for moderately conditioned problems or with moderate dynamic range in order to 
be accurate, reproducible, but also high performing since the ExBLAS version can be very resource demanding. To sum up, if the information about the problem is know in advance, it is good to explore the FPE-based implementation.
%

\section{Experimental Results}
\label{sec:results}

\subsection{Setup}
\label{sec:setup}
The experiments in this section employed IEEE754 double-precision arithmetic and
were carried out in two different clusters:
\begin{itemize}
    \vspace{-4pt}
    \setlength\itemsep{-1.5pt}
    \item The {\em MareNostrum4} (MN4) supercomputer at {\em Barcelona Supercomputing Center (BSC)}: This platform consists of SD530 Compute Racks with an Intel Omni-Path high performance network interconnect.
    Each node comprises two 24-core Intel Xeon Platinum 8160 processors (2.10~GHz) and 96~Gbytes of DDR4 RAM. 
    The platform runs the SuSE Linux Enterprise Server operating system.
    The codes in this platform were compiled using GCC v7.2.0, Intel MPI v2018.1, and MKL v2017.4. 
    \item The {\em Tintorrum} cluster at \textit{Universitat Jaume~I}: 
    This is a 8-node cluster, where each node is equipped with
    two 8-core Intel Xeon(R) E5-2630v3 processors (Haswell-EP) 
    (for a total of 128 cores), running at 2.4~GHz, with 20~MBytes of L3 on-chip cache (LLC or last level of cache), and 
    with 64~GBytes of DDR3 RAM. The operating system running in the cluster is Linux version 2.6.32-642.4.2.el6.centos.plus.x86\_64.
     The codes were compiled with GCC v5.3.0, OpenMPI v1.10.2, and Intel MKL v2017.1.    
    \vspace{-4pt} 
\end{itemize}

For the experimental analysis, we leveraged a sparse
s.p.d. coefficient matrix arising from the finite-difference method of a 3D Poisson's equation with 27 stencil points. 
The fact that the vector involved in the \spmv kernel has to be replicated in all MPI ranks constrains 
the size of the largest problem that can be solved. 
Given that the theoretical cost of PCG is $t_c \approx 2nnz+ 7n$ floating-point arithmetic operations, 
where $nnz$ denotes  the number of nonzeros of the original matrix and its size $n$, 
the execution time of the method is usually dominated by that of the \spmv kernel.
Therefore, in order to analyze the weak scalability of the method, we maintain the number of non-zero entries per node. 
For this purpose, 
we modified the original matrix, transforming it into a band matrix, 
where the lower and upper bandwidths (\textit{bandL} and \textit{bandU}, respectively) 
depend on the number of nodes employed in the experiment as follows:
\begin{multline*}
bandL = bandU = 100 \times \#nodes \quad \Longrightarrow\\
\quad nnz = ( bandL + bandU + 1 ) \times n.
\end{multline*}
With 8 nodes in Tintorrum and 16 in MN4, the bandwidth 
ranges between 100 and 800 in the first platform, and from 100 to 1,600 in the second one. 
With this approach we can then maintain the number of rows/columns of the matrix equal to $n$=4,000,000, while 
increasing its bandwidth and, therefore, the computational workload
proportionally to the hardware resources, as required
in a weak scaling experiment. 

The right-hand side vector $b$ in the iterative solvers was always initialized to the product of $A$ with a vector
containing ones only; 
and the PCG iteration was started with the initial guess $x_0=0$. 
The parameter that controls the convergence of the iterative process was set to $10^{-8}$.

\subsection{Performance Evaluation}
We analyze the performance of two reproducible versions of the PCG algorithm parallelized with MPI: one that relies on the ExBLAS approach, and an alternative variant that is based on floating-point expansions (FPEs) of size eight with the early-exit technique. Hereafter, we will refer to them as {\em Exblas} and {\em Opt} (or {\em FPE8EE}), accordingly. Our experiments evaluate the strong and weak scaling of these reproducible implementations compared against the regular (non-deterministic) version of PCG; all three versions are implemented with MPI + OpenMP tasks.


We next analyze the performance of the three implementations in the aforementioned clusters. On the one hand, in order to assess the strong scalability, we fix the matrix size to $n$=16,000,000 and the size of the upper and lower bandwidth to 100, as we increase the number of cores. On the other hand, in order to analyze the weak scalability, we proceed as explained earlier, fixing the matrix size to $n$=4,000,000 and increasing the bandwidth from 100 to $100 \times $mnodes (with mnodes=16 in MN4 and mnodes=8 in Tintorrum).

\begin{table*}
\begin{center}
\begin{tabular}{|c||c|c|c||c|c|c|}
   \hline
   \multicolumn{7}{|c|}{Execution time in seconds of the implementations in MN4} \\ \hline
   Number  & \multicolumn{3}{c||}{Weak scaling} & \multicolumn{3}{c|}{Strong scaling} \\ \cline{2-4}\cline{5-7}
     of cores & {\em Regular} & {\em Exblas} & {\em Opt}  & {\em Regular} & {\em Exblas} & {\em Opt}    \\ \hline  \hline
  48  &  3.5349E+00 & 8.8568E+00 & 7.7153E+00 & 1.3280E+01 & 3.5312E+01 & 2.9730E+01  \\
  96  &  3.1697E+00 & 5.9492E+00 & 5.4720E+00 & 7.6761E+00 & 1.8550E+01 & 1.6142E+01  \\
  192 &  2.9610E+00 & 4.7935E+00 & 4.5801E+00 & 5.1802E+00 & 1.0523E+01 & 9.3799E+00  \\
  384 &  2.8018E+00	& 3.9885E+00 & 3.8810E+00 & 3.9321E+00 & 6.5620E+00 & 6.0571E+00  \\
  768 &  3.5905E+00 & 4.7965E+00 & 4.7348E+00 & 3.6662E+00 & 5.0488E+00 & 4.7846E+00  \\ \hline 
   \multicolumn{7}{|c|}{Execution time in seconds of the implementations on Tintorrum} \\ \hline
   Number  & \multicolumn{3}{c||}{Weak scaling} & \multicolumn{3}{c|}{Strong scaling} \\ \cline{2-4}\cline{5-7}
   of cores & {\em Regular} & {\em Exblas} & {\em Opt}  & {\em Regular} & {\em Exblas} & {\em Opt}    \\ \hline  \hline
    16 &  8.3203E+00 &	1.4222E+01	& 1.3014E+01 & 3.2747E+01	& 5.7285E+01	& 5.1238E+01 \\
    32 &  1.6787E+01 &	2.2833E+01	& 2.1898E+01 & 4.8481E+01	& 7.0335E+01	& 6.8607E+01\\
    64 &  1.8877E+01 &	2.1114E+01	& 2.0992E+01 & 5.8668E+01	& 7.2928E+01	& 7.1930E+01\\
    128 & 1.8322E+01 &	2.0331E+01	& 2.0156E+01 & 6.4591E+01	& 6.8174E+01	& 6.7651E+01 \\
    \hline
\end{tabular}
\end{center}
\caption {\label{tab:comparison} Timings of different implementations of the Preconditioned Conjugate Gradient method in MN4 and Tintorrum.}
\end{table*}
\Cref{tab:comparison} reports the total execution time (averaged for 5 different executions) of the different MPI + OpenMP tasks PCG solvers on both platforms, varying the number of cores (from 48 to 768 in MN4 and from 16 to 128 in Tintorrum) as we maintain the problem size.  
We tested different computations of MPI processes per node and OpenMP threads per process: the best performing in MN4 is 8 MPI process with 6 OpenMP threads each, and the optimum combination on Tintorrum is 8 MPI process with 2 OpenMP threads each. The weak scaling experiment offers notable results, as, when executing the algorithms in more than one node (up to 48 cores in MN4 and up to 16 cores on Tintorrum) while increasing proportionally the problem, the execution time is maintained. The executions on one node show a different behavior because the communication is in general faster as it entails no inter-node communication. Notably, these extra (local) operations of both ExBLAS and Opt implementations have a positive effect on scalability on the larger node count due to better ratio of computations to communication compared with the original version.
The behaviour of the strong scaling experiment could be expected for a parallel algorithm dealing with a sparse linear algebra operation. This experiment in particular reports an important 
increase of the overhead when the number of nodes becomes large as the communication cost then dominates the execution time. But, the overhead of the reproducible versions decreases due to the favorable ratio between computations and communication. Unfortunately, we cannot evaluate a larger problem to increase the weight of the computational cost, as the problem dimension is constrained by the node
memory capacity.

\begin{figure*}[th!]
\begin{center}
\begin{tabular}{cc}
\begin{minipage}[t]{0.48\textwidth}
\includegraphics[width=\textwidth,height=6.2cm]{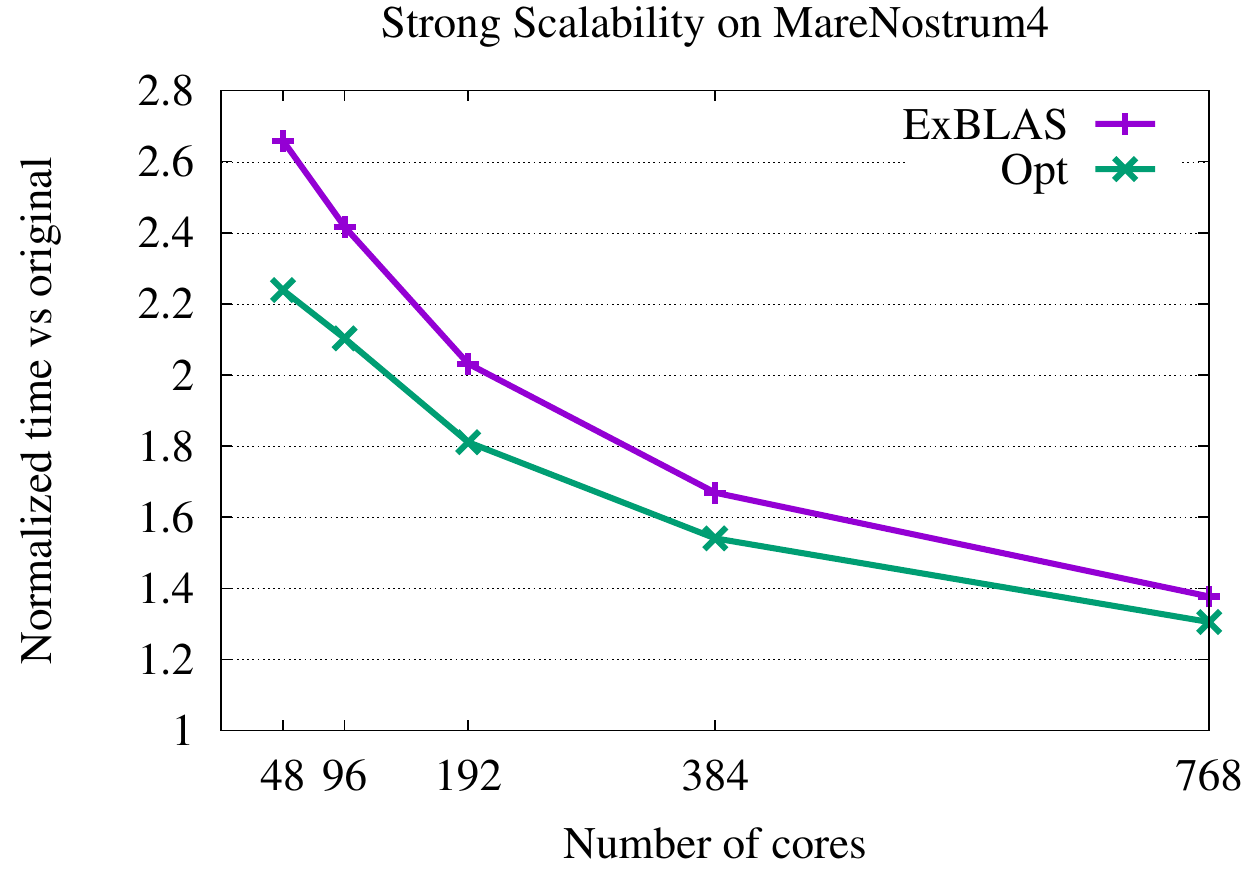}
\end{minipage}
&
\begin{minipage}[t]{0.48\textwidth}
\includegraphics[width=\textwidth,height=6.2cm]{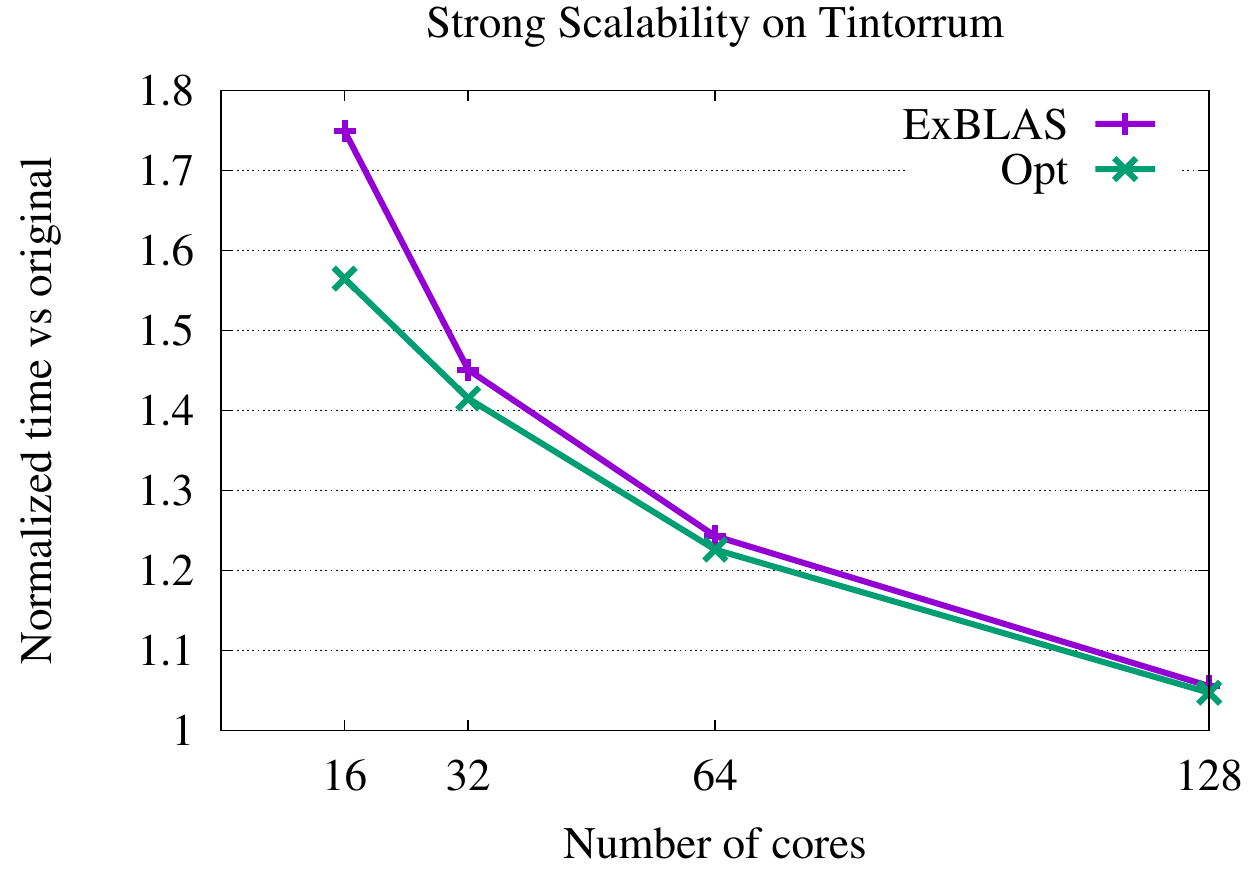}
\end{minipage}
\\
\begin{minipage}[t]{0.48\textwidth}
\includegraphics[width=\textwidth,height=6.2cm]{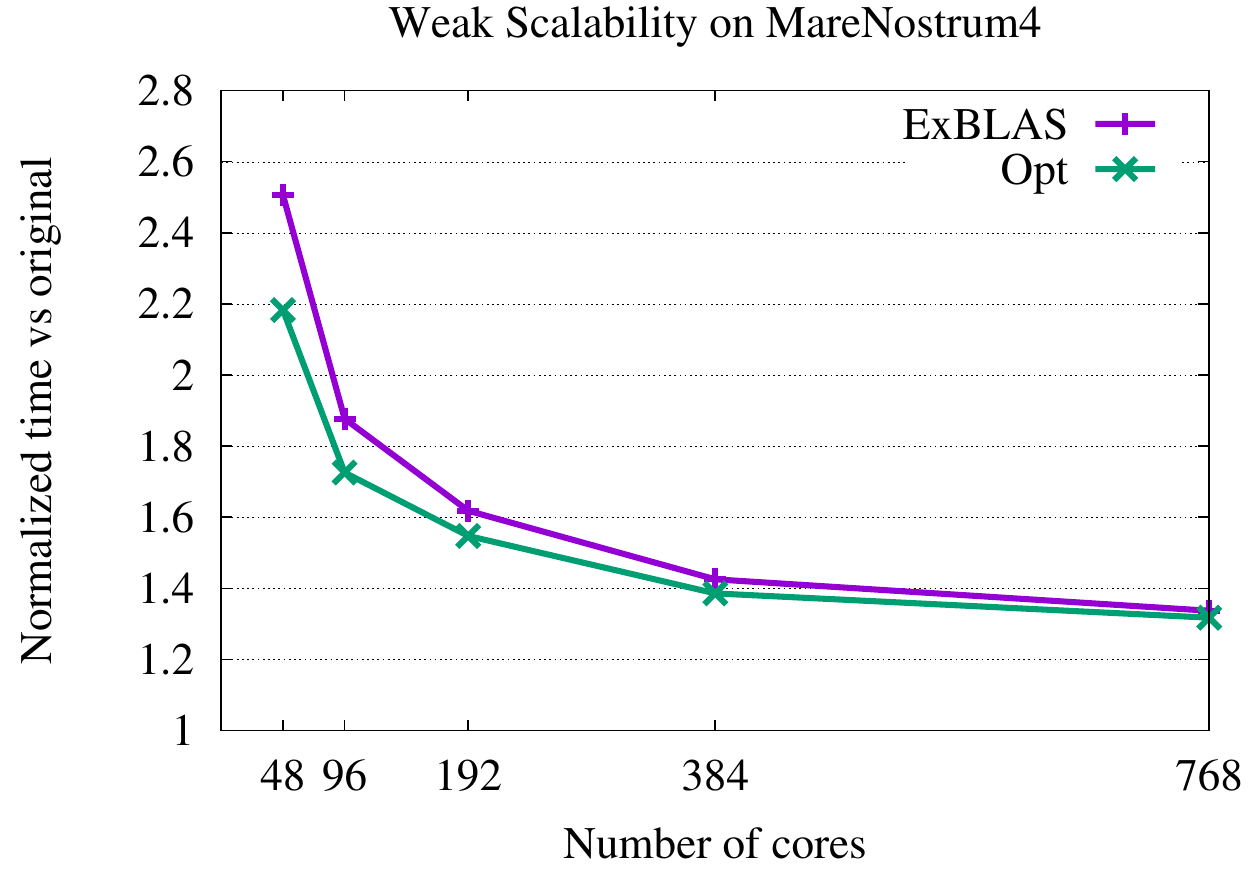}
\end{minipage}
&
\begin{minipage}[t]{0.48\textwidth}
\includegraphics[width=\textwidth,height=6.2cm]{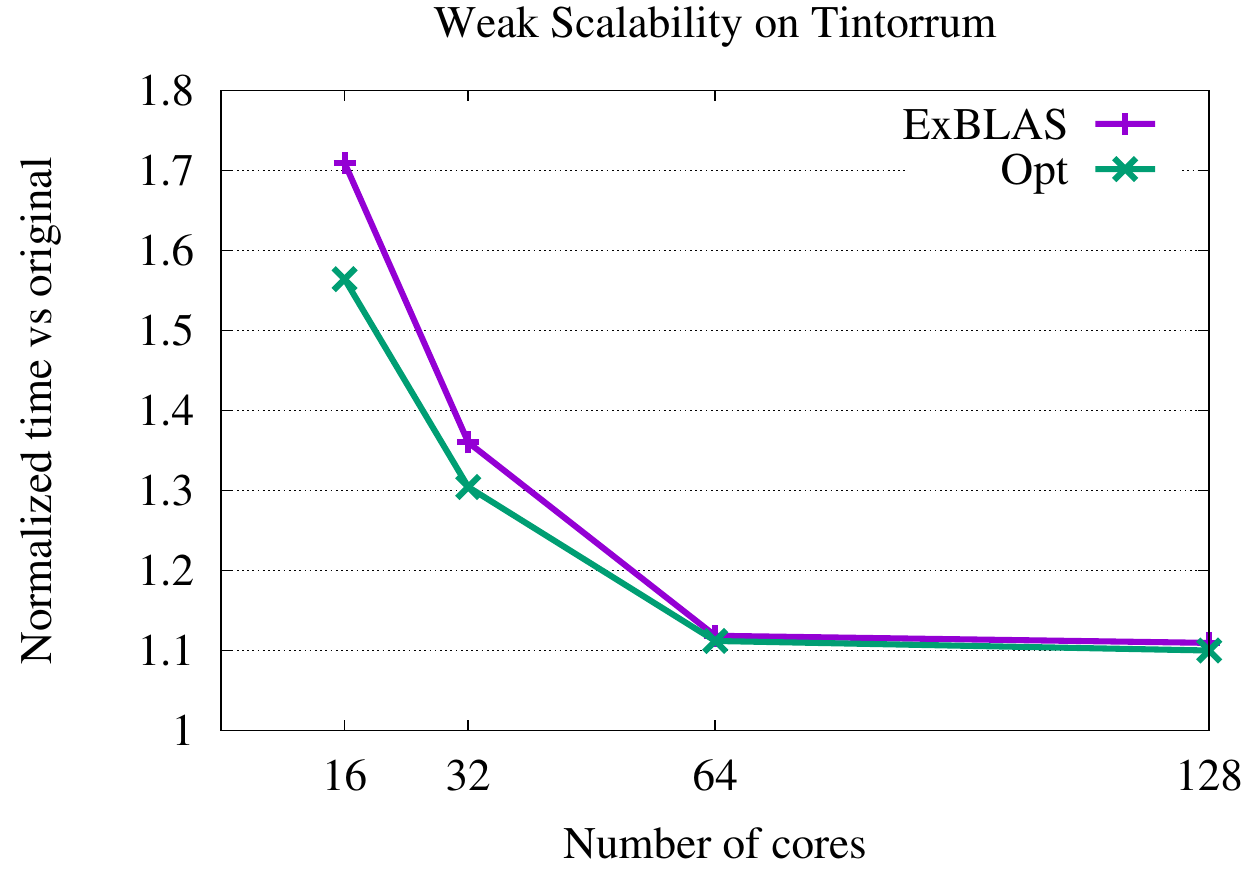}\\
\end{minipage}
\end{tabular}
\vspace{-10pt}
\caption{\label{fig:performance} Analysis of the strong (top) and weak (bottom) scalability of the two reproducible versions of the MPI + OpenMP tasks PCG; the time is normalized with respect to the regular non-deterministic MPI version.}
\end{center}
\end{figure*}

\begin{table*}
\begin{center}
\resizebox{\textwidth}{!}{%
\begin{tabular}{|c|l|l|l|l|}
   \hline
   Iteration  & \multicolumn{4}{c|}{Residual} \\ 
   \cline{2-5}
   & {\em MPFR} & {\em Original} 1 core & {\em Original} 48 cores & {\em Exblas \& FPE8EE} \\ \hline  \hline
   0 & 0x1.19f179eb7f032p+49 & 0x1.19f179eb7f03{\bf 3}p+49 & 0x1.19f179eb7f03{\bf 3}p+49 & 0x1.19f179eb7f032p+49\\
   2 & 0x1.f86089ece9f75p+38 & 0x1.f86089ece{\bf 5bd4}p+38 & 0x1.f86089ece{\bf af76}p+38 & 0x1.f86089ece9f75p+38\\
   9 & 0x1.fc59a29d329ffp+28 & 0x1.fc59a29d3{\bf 599a}p+28 & 0x1.fc59a29d32{\bf d1b}p+28 & 0x1.fc59a29d329ffp+28\\
   10 & 0x1.74f5ccc211471p+22 & 0x1.74f5ccc{\bf 1d03cb}p+22 & 0x1.74f5ccc2{\bf 01246}p+22 & 0x1.74f5ccc211471p+22\\
    ... & ... & ... & ...& ...\\
   40 & 0x1.7031058eb2e3ep-19 & 0x1.7031058{\bf dd6bcf}p-19 & 0x1.7031058e{\bf af4c2}p-19 & 0x1.7031058eb2e3ep-19\\
   42 & 0x1.4828f76bd68afp-23 & 0x1.4828f76{\bf d1aa3}p-23 & 0x1.4828f76bd{\bf a71a}p-23 & 0x1.4828f76bd68afp-23\\
   45 & 0x1.8646260a70678p-26 & 0x1.8646260a{\bf 2dae8}p-26 & 0x1.8646260a{\bf 6da06}p-26 & 0x1.8646260a70678p-26\\
   47 & 0x1.13fa97e2419c7p-33 & 0x1.13fa97e{\bf 1e76bf}p-33 & 0x1.13fa97e24{\bf 0f7c}p-33 & 0x1.13fa97e2419c7p-33\\   
   \hline
\end{tabular}
}
\end{center}
\caption {\label{tab:acc} Accuracy and reproducibility comparison on the intermediate and final residual against MPFR for a matrix with the condition number of $10^{12}$. The matrix is generated following the procedure from~\Cref{sec:setup} with n=4,019,679 ($159^3$) and the bandwidth of size 200.}
\end{table*}

Figure \ref{fig:performance} reports the total execution time (averaged for 5 different executions) of the reproducible MPI PCG solvers for the two clusters normalized with respect to the execution time of the regular MPI version, when we vary the number of cores (from 48 to 768 in MN4 and from 16 to 128 in Tintorrum).
Specifically, in the two top plots we present the strong scaling evaluation. In these graphs, we can observe that the difference of both versions with respect to the regular one is higher on a small number of cores, and it decreases with the core count. 
We observe that the overhead of both the {\em Exblas}  and {\em Opt} implementations compared with the regular version is smooth and decreasing: from 2.66x and 2.24x on the single node to 37.7\,\% and 30.5\,\% on 16 nodes on MN4 for {\em Exblas}  and {\em Opt}, respectively; and, on Tintorrum from 74.9\,\% and 56.5\,\% on the single node to 5.6\,\% and 4.7\,\% on 8 nodes for {\em Exblas}  and {\em Opt}, accordingly. 
Moreover, the overhead between  {\em Exblas}  and {\em Opt} versions decreases, e.g. from 18\,\% to 6\,\% in MN4, on the large core count: this is due to very similar implementations of both since {\em Exblas} underneath relies upon FPE8EE for the OpenMP \dotp\ products. Note that such difference is much larger for the pure MPI implementation~\cite{iakymchuk19jcam}.

The two bottom graphs in Figure~\ref{fig:performance} expose the weak scaling evaluation, where we set the number of non-zeros of the sparse matrix to be roughly proportional to the number of cores, increasing the size of the band of the matrix, as discussed in~\Cref{sec:setup}. These results show that both versions offer similar performance to the baseline on the large number of cores. For instance, the overheads are 33.70\,\% and 31.75\,\% for the {\em Exblas} and {\em Opt} implementations in MN4, respectively, and only 11\,\% and 10\,\% for {\em Exblas} and {\em Opt} on Tintorrum, accordingly. As in the strong scaling analysis, the {\em Opt} version outperforms the {\em Exblas} implementation. If we compare the results in both clusters, we can observe that they are more stable in Tintorrum because the number of cores per node is smaller in this platform than in MN4. 


\subsection{Accuracy and Reproducibility Evaluation}
In addition to the performance results, we report also the results of the accuracy and reproducibility evaluation. For that, we develop a generator of ill-conditioned matrices. This generator scales the first row and the first column of the matrix so that the \dotp\ product determines the condition number of the matrix. Additionally, we derive a sequential version of the code that relies on the GNU Multiple Precision Floating-Point Reliably (MPFR) library~\cite{MPFR} -- a C library for multiple (arbitrary) precision floating-point computations on CPUs -- as a highly accurate reference implementation. This implementation uses 2,048 bits of accuracy for computing the \dotp\ product (192 bits for internal product of two floating-point numbers) and performs correct rounding of the computed result to double precision.
 
\Cref{tab:acc} reports the intermediate and final residual on each iteration of the PCG solver for the matrix with the number of rows/columns equal to n=4,019,679 ($159^3$), the bandwidth of size 200, and the condition number of $10^{12}$. The results are exposed with all digits in hexadecimal. For this test, the tolerance was set to $10^{-8}$ and it took $47$ iterations for all four implementations to converge under this accuracy requirement.
We used one node of MN4 with $48$ processes each pinned to one core. We present only few iterations, but the difference is present in all iterations. The {\em ExBLAS} and {\em Opt} implementations deliver both accurate and reproducible results that are identical with the MPFR library. Note that these results are identical to the ones from the pure MPI implementations in~\cite{iakymchuk19jcam} and only the results of the original code differ. The original code shows the difference from one digit on the initial iteration and up to five digits on the 45th iteration on 48 cores (8 MPI processes with 6 OpenMP threads per each). We also add the results of the original code on one core/process to highlight the reproducibility issue. To show these results, we merge the two  columns of the ExBLAS and Opt results as they are identical. 

We assume that this discrepancy in accuracy and reproducibility becomes larger at scale (more nodes) due to the stronger impact of the topology and reduction trees. 

\subsection{Evaluation using the SuiteSparse matrix collection}
We conduct a set of tests using real cases from the SuiteSparse matrix collection. We select matrices with various condition numbers starting from $1.49e+04$ up to $2.22e+18$\footnote{In practice, selecting coefficient matrices for the linear systems for which $cond(A)\leq1e+12$ would have been more realistic due to the limits of the double precision arithmetic.}, with as many as one million nonzero elements. \Cref{tab:suitesparse:mn4} presents our experimental results on a single node using eight MPI processes and six OpenMP threads per process on the MareNostrum4 cluster. For each test matrix, we report the number of iterations required to reach the tolerance of $10^{-8}$ for residual, the direct error computed as in Section 3.5.1~\cite{GVL:2013}, and the total execution time. We have selected the direct error instead of residual since it is known that smaller residual does not imply higher accuracy, meaning solutions with smaller residual might be less accurate. This is confirmed for the first three matrices for which the residual suffices the tolerance but the direct error is still large. The direct error as well as the number of iterations are identical for both {\em ExBLAS} and {\em Opt} variants, hence we merge these columns. Our reproducible variants require a smaller number of iterations than the original version for the msc01050, olafu, and bcsstk28 matrices. For instance, for the olafu matrix (1,015,156 nonzero elements and condition number $7.61e+11$), both ExBLAS and Opt variants require 42,342 iterations, while the original version needs 43,046 iterations. For a few cases, our reproducible variants may perform slightly more iterations than the non-reproducible variants due to the differences in the accumulation of rounding errors arising form distinct optimizations in the codes. The overhead of our reproducible variants can be as low as $32.5$\,\% for the 494\_bus matrix but can reach $3.46\times$ for the sts4098 matrix. This overhead is expected and is inline with the pattern from~\Cref{fig:performance}, where the largest overhead is observed on a single node. Moreover, we also run tests using eight MPI processes and two OpenMP threads per process --  the {\em Opt} and {\em ExBLAS} results are again identical in terms of the number of iterations, residuals, and direct errors.
 
Furthermore, we conduct similar experiments on the Tintorrum cluster. \Cref{tab:suitesparse} presents the results of these experiments. There, we also use one node but four MPI processes and four OpenMP threads per process. These results show a similar trend to that of MareNostrum4: smaller number of iterations of the {\em ExBLAS} and {\em Opt} versions for plat1919, olafu, gyro\_k, bcsstk28, and msc04515 matrices; the overhead of reproducible versions as small as $88.1$\,\% for the 494\_bus matrix but may also grow up to few times. Notably, both {\em ExBLAS} and {\em Opt} versions deliver identical results, excluding timings, on the MareNostrum4 and Tintorrum clusters.

In addition, we conduct experiments 
using the pure MPI versions of the Reproducible Preconditioned Conjugate Gradient~\cite{iakymchuk19jcam} on the MareNostrum4 and Tintorrum clusters, see~\Cref{tab:suitesparse:purempi:mn4,tab:suitesparse:purempi}. 
We observe that the number of iterations, residuals, direct errors, the final error, and vector-solutions are identical to those produced by the MPI+OpenMP tasks versions. Hence, our reproducible strategies ensure {\em cross-cluster reproducibility of PCG} implemented with pure MPI as well as the hybrid MPI + OpenMP tasks models.
 
\begin{table*}
\begin{center}
\resizebox{\textwidth}{!}{
\begin{tabular}{|l|r|r|r|r|l|l|l|l|l|}
   \hline
   Matrix  & Nonzeros & $cond(A)$ & \multicolumn{2}{c|}{Iterations}& \multicolumn{2}{c|}{Direct error} & \multicolumn{3}{c|}{Time [secs]} \\ \cline{4-10}
   & & & {\em Orig} & {\em Opt{\&}ExBLAS} & {\em Orig} & {\em Opt{\&}ExBLAS} & {\em Orig} & {\em Opt} & {\em Exblas}    \\ \hline  \hline
  plat1919 & 32,399 & 2.22e+18 & 28200 & 28347 & 0x1.d1fd2948ac992p+4 &  0x1.d1fd1980ddc3dp+4 & 2.89e+00 & 5.79e+00 & 6.74e+00 \\
  msc01050 & 26,198 & 9.00e+15 & 1459 & 1441 & 0x1.fe62a1a8f70acp-7 & 0x1.c06963286be9ap-7 & 1.35e-01 & 2.27e-01 & 2.48e-01 \\
  mhd4800b & 27.520 & 1.03e+14 & 32 & 32 & 0x1.171f2d2a7c15dp-7 & 0x1.171f2cf90e554p-7 & 3.76e-03 & 1.10e-02 & 1.30e-02  \\ 
  olafu & 1,015,156 & 7.61e+11 & 43046 & 42342 &  0x1.4683bc1ddab86p-28 & 0x1.3c603001a3c8fp-28 & 3.54e+01 & 6.84e+01 & 7.60e+01 \\
  gyro\_k & 1,021,159 & 1.10e+09 & 16064 & 16075 & 0x1.7f300f81c65c7p-26 & 0x1.7ef8863fde778p-26 & 1.41e+01 & 2.78e+01 & 3.10e+01\\
  bcsstk28 & 219,024 & 6.28e+09 & 5592 & 5483 & 0x1.dd4e900472e3dp-31 & 0x1.ba21beeb93c43p-31 & 1.25e+00 & 2.51e+00 & 2.85e+00 \\
  bcsstk13 & 83,883 & 5.64e+08 & 2571 & 2571 & 0x1.03a9e5339d79fp-34 & 0x1.5e81334a6997bp-34 & 3.66e-01 & 6.47e-01 & 7.37e-01\\
  sts4098 & 72,356 & 3.56e+07 & 666 & 668 & 0x1.ad954060aba53p-38 & 0x1.5a2eda56201fbp-38 & 8.87e-02 & 2.28e-01 & 2.61e-01\\
  494\_bus & 1,666 & 3.89e+06 & 410 & 410 & 0x1.22befca9188e3p-35 & 0x1.b83293969f70bp-36 & 3.29e-02 & 4.36e-02 & 5.05e-02\\
  msc04515 & 97,707 & 4.78e+05 & 4883 & 4885 & 0x1.41d64ef6a77bfp-32 & 0x1.4b0c4d82346dap-32 & 6.85e-01 & 1.80e+00 & 2.07e+00\\
  bcsstk27 & 56,126 & 1.49e+04 & 331 & 331 & 0x1.3f45a221626fdp-40 & 0x1.2b65be5099a69p-40 & 3.66e-02 & 5.94e-02 & 6.60e-02\\ 
  
 \hline 
\end{tabular}
}
\end{center}
\caption {\label{tab:suitesparse:mn4} Evaluation of different {\em MPI + OpenMP tasks} implementations of the PCG method using test matrices from the SuiteSparse matrix collection on one {\em MareNostrum4's} node with eight MPI processes and six OpenMP threads per process.}
\end{table*}

\begin{table*}
\begin{center}
\resizebox{\textwidth}{!}{
\begin{tabular}{|l|r|r|r|r|l|l|l|l|l|}
   \hline
   Matrix  & Nonzeros & $cond(A)$ & \multicolumn{2}{c|}{Iterations}& \multicolumn{2}{c|}{Direct error} & \multicolumn{3}{c|}{Time [secs]} \\ \cline{4-10}
   & & & {\em Orig} & {\em Opt{\&}ExBLAS} & {\em Orig} & {\em Opt{\&}ExBLAS} & {\em Orig} & {\em Opt} & {\em Exblas}    \\ \hline  \hline
  plat1919 & 32,399 & 2.22e+18 & 29133 & 28347 & 0x1.d1fccb48b0708p+4 & 0x1.d1fd1980ddc3dp+4 & 2.21e+00 & 6.07e+00 & 7.65e+00  \\
  msc01050 & 26,198 & 9.00e+15 & 1441 & 1441 & 0x1.20ef1ec5aba0fp-6 & 0x1.c06963286be9ap-7 &  1.07e-01 & 2.26e-01 & 2.74e-01  \\
  mhd4800b & 27,520 & 1.03e+14 & 32 & 32 & 0x1.171f2d405896bp-7 & 0x1.171f2cf90e554p-7 &  4.35e-02 & 1.06e-01 & 8.98e-02   \\ 
  olafu & 1,015,156 & 7.61e+11 & 44309 & 42342& 0x1.580f68bcf7c59p-28 & 0x1.3c603001a3c8fp-28 &  5.80e+01 & 1.10e+02  & 1.28e+02 \\
  gyro\_k & 1,021,159 & 1.10e+09 & 16623 & 16075 & 0x1.70c410f76c1f3p-26 & 0x1.7ef8863fde778p-26 &  2.12e+01 & 4.24e+01 & 4.98e+01 \\
  bcsstk28 & 219,024 & 6.28e+09 & 5485 & 5483 & 0x1.b33c0d0a65819p-31 & 0x1.ba21beeb93c43p-31 &  1.59e+00 & 3.50e+00  & 4.17e+00 \\
  bcsstk13 & 83,883 & 5.64e+08 & 2554 & 2571 & 0x1.1f45539d3ad76p-34 & 0x1.5e81334a6997bp-34 & 3.87e-01 & 8.04e-01 & 9.58e-01  \\
  sts4098 & 72,356 & 3.56e+07 & 666 & 668 & 0x1.75b26e5cc575ep-38 & 0x1.5a2eda56201fbp-38 & 1.02e-01 & 3.00e-01 & 3.69e-01 \\
  494\_bus & 1,666 & 3.89e+06 & 409 & 410 & 0x1.8a70c6145af0bp-33 & 0x1.b83293969f70bp-36 & 1.76e-02 & 3.31e-02  & 4.14e-02 \\
  msc04515 & 97,707 & 4.78e+05 & 5138 & 4885 & 0x1.318ee7cc28729p-32 & 0x1.4b0c4d82346dap-32 & 8.34e-01 & 2.43e+00 & 2.98e+00  \\
  bcsstk27 & 56,126 & 1.49e+04 & 331 & 331 & 0x1.287d86ae5b307p-40 & 0x1.2b65be5099a69p-40 & 3.34e-02 & 6.53e-02 & 7.81e-02  \\ 
 \hline 
\end{tabular}
}
\end{center}
\caption {\label{tab:suitesparse} Evaluation of different {\em MPI + OpenMP tasks} implementations of the PCG method using test matrices from the SuiteSparse matrix collection on one {\em Tintorrum's} node with four MPI processes and four OpenMP threads per process.}
\end{table*}

\begin{table*}
\begin{center}
\resizebox{\textwidth}{!}{
\begin{tabular}{|l|r|r|r|r|l|l|l|l|l|}
   \hline
   Matrix  & Nonzeros & $cond(A)$ & \multicolumn{2}{c|}{Iterations}& \multicolumn{2}{c|}{Direct error} & \multicolumn{3}{c|}{Time [secs]} \\ \cline{4-10}
   & & & {\em Orig} & {\em Opt{\&}ExBLAS} & {\em Orig} & {\em Opt{\&}ExBLAS} & {\em Orig} & {\em Opt} & {\em Exblas}    \\ \hline  \hline
  plat1919 & 32,399 & 2.22e+18 & 28404 & 28347 &  0x1.d1fd13459efb2p+4 &  0x1.d1fd1980ddc3dp+4 & 2.09e+00 & 7.25e+00 & 1.15e+01 \\
  msc01050 & 26,198 & 9.00e+15 & 1449 & 1441 & 0x1.1e960e3dd96adp-6 & 0x1.c06963286be9ap-7 & 1.18e-01 & 3.51e-01 & 4.95e-01 \\
  mhd4800b & 27,520 & 1.03e+14 & 32 & 32 & 0x1.171f2d513bf1fp-7 & 0x1.171f2cf90e554p-7 & 5.17e-03 & 9.00e-03 & 2.16e-02  \\ 
  olafu & 1,015,156 & 7.61e+11 & 44872 & 42342 & 0x1.284b2460347acp-28 &  0x1.3c603001a3c8fp-28 & 1.38e+01 & 2.22e+01 & 7.77e+01 \\
  gyro\_k & 1,021,159 & 1.10e+09 & 16577 & 16075 & 0x1.76957c0ecf952p-26 & 0x1.7ef8863fde778p-26 & 5.29e+00 & 8.71e+00 & 3.14e+01\\
  bcsstk28 & 219,024 & 6.28e+09 & 5736 & 5483 &  0x1.e9c64a28a93dfp-31 & 0x1.ba21beeb93c43p-31 & 8.43e-01 & 1.70e+00 & 3.69e+00 \\
  bcsstk13 & 83,883 & 5.64e+08 & 2571 & 2571 & 0x1.03a9e5339d79fp-34 & 0x1.5e81334a6997bp-34 & 2.32e-01 & 5.07e-01 & 3.01e+00\\
  sts4098 & 72,356 & 3.56e+07 & 666 & 668 & 0x1.ad954060aba53p-38 & 0x1.5a2eda56201fbp-38 & 6.44e-02 & 1.85e-01 & 1.47e+00\\
  494\_bus & 1,666 & 3.89e+06 & 410 & 410 & 0x1.98d21a409a23cp-36 & 0x1.b83293969f70bp-36 & 4.47e-02 & 9.29e-02 & 1.20e-01\\
  msc04515 & 97,707 & 4.78e+05 & 4883 & 4885 & 0x1.41d64ef6a77bfp-32 & 0x1.4b0c4d82346dap-32 & 4.89e-01 & 1.47e+00 & 1.18e+01\\
  bcsstk27 & 56,126 & 1.49e+04 & 331 & 331 & 0x1.3f45a221626fdp-40 & 0x1.2b65be5099a69p-40 & 2.02e-02 & 4.50e-02 & 1.22e-01\\   
 \hline 
\end{tabular}
}
\end{center}
\caption {\label{tab:suitesparse:purempi:mn4} 
Evaluation of different {\em pure MPI} implementations~\cite{iakymchuk19jcam} of the PCG method using test matrices from the SuiteSparse matrix collection on one {\em MareNostrum4's} node with 48 MPI processes.}
\end{table*}

\begin{table*}
\begin{center}
\resizebox{\textwidth}{!}{
\begin{tabular}{|l|r|r|r|r|l|l|l|l|l|}
   \hline
   Matrix  & Nonzeros & $cond(A)$ & \multicolumn{2}{c|}{Iterations}& \multicolumn{2}{c|}{Direct error} & \multicolumn{3}{c|}{Time [secs]} \\ \cline{4-10}
   & & & {\em Orig} & {\em Opt{\&}ExBLAS} & {\em Orig} & {\em Opt{\&}ExBLAS} & {\em Orig} & {\em Opt} & {\em Exblas}    \\ \hline  \hline
  plat1919 & 32,399 & 2.22e+18 & 28225 & 28347 &  0x1.d1fd2b1f22f89p+4 &  0x1.d1fd1980ddc3dp+4 & 1.05e+00 & 2.27e+00 & 1.21e+01 \\
  msc01050 & 26,198 & 9.00e+15 & 1440 & 1441 & 0x1.91eb8c4cf549ep-7 & 0x1.c06963286be9ap-7 & 5.85e-02 & 1.01e-01 & 4.14e-01 \\
  mhd4800b & 27,520 & 1.03e+14 & 32 & 32 & 0x1.171f2d071bbecp-7 & 0x1.171f2cf90e554p-7 & 2.05e-03 & 4.82e-03 & 3.28e-02  \\ 
  olafu & 1,015,156 & 7.61e+11 & 44840 & 42342 &  0x1.73ee2c0ee4f91p-28 &  0x1.3c603001a3c8fp-28 & 2.08e+01 & 3.39e+01 & 1.55e+02 \\
  gyro\_k & 1,021,159 & 1.10e+09 & 16518 & 16075 &  0x1.7c04191d8b8d9p-26 & 0x1.7ef8863fde778p-26 & 8.31e+00 & 1.38e+01 & 6.28e+01\\
  bcsstk28 & 219,024 & 6.28e+09 & 5640 & 5483 &  0x1.d8e49bef6a637p-31 & 0x1.ba21beeb93c43p-31 & 5.58e-01 & 1.06e+00 & 5.46e+00 \\
  bcsstk13 & 83,883 & 5.64e+08 & 2337 & 2571 &  0x1.040b38e017aeep-34 & 0x1.5e81334a6997bp-34 & 1.36e-01 & 2.68e-01 & 1.23e+00\\
  sts4098 & 72,356 & 3.56e+07 & 668 & 668 & 0x1.77b31324422c1p-38 & 0x1.5a2eda56201fbp-38 & 4.36e-02 & 9.92e-02 & 6.02e-01\\
  494\_bus & 1,666 & 3.89e+06 & 410 & 410 &  0x1.f52f8f4c274dp-37 & 0x1.b83293969f70bp-36 & 1.21e-02 & 1.63e-02 & 5.78e-02\\
  msc04515 & 97,707 & 4.78e+05 & 4874 & 4885 &  0x1.2dfe5e95c1703p-32 & 0x1.4b0c4d82346dap-32 & 3.26e-01 & 7.56e-01 & 4.80e+00\\
  bcsstk27 & 56,126 & 1.49e+04 & 331 & 331 & 0x1.2c11a8f939c1dp-40 & 0x1.2b65be5099a69p-40 & 1.45e-02 & 2.53e-02 & 1.04e-01\\   
 \hline 
\end{tabular}
}
\end{center}
\caption {\label{tab:suitesparse:purempi} Evaluation of different {\em pure MPI} implementations~\cite{iakymchuk19jcam} of the PCG method using test matrices from the SuiteSparse matrix collection on one {\em Tintorrum's} node with 16 MPI processes.}
\end{table*}

\section{Related Work}
\label{sec:related:works}
To enhance reproducibility, Intel proposed the ``Conditional Numerical Reproducibility'' (CNR) option in its Math Kernel Library (MKL).
Although CNR guarantees reproducibility, it does not ensure correct rounding, meaning the accuracy is arguable. 
Additionally, the cost of obtaining reproducible results with CNR is high. For instance, for large arrays 
the MKL's summation with CNR was almost 2x slower than the regular MKL's summation on the Mesu cluster hosted at the Sorbonne University~\cite{Collange15Parco}.

Demmel and Nguyen implemented a family of algorithms -- that originate from the works by Rump, Ogita, and Oishi~\cite{RuOgOi2010,RuOgOi2008b,RuOgOi2008a} -- for reproducible summation in floating-point arithmetic~\cite{Demmel13Arith0,Demmel14OneRed}. These algorithms always return the same answer. They first compute an absolute bound of the sum and then round all numbers to a fraction of this bound. In consequence, the addition of the rounded quantities is exact, however the computed sum using their implementations with two or three bins is not correctly rounded. 
Their results yielded roughly $20$\,\% overhead on $1024$ processors (CPUs only) compared to the Intel MKL \texttt{dasum()}, but it shows $3.4$ times 
slowdown on $32$ processors (one node). 
Ahrens, Nguyen, and Demmel extended their concept to few other reproducible BLAS routines, distributed as the ReproBLAS 
library\footnote{\url{http://bebop.cs.berkeley.edu/reproblas/}}, but only with parallel reproducible reduction. Furthermore, the ReproBLAS effort was extended to reproducible  tall-skinny QR~\cite{NguyenD15}.

The other approach to ensure reproducibility is called ExBLAS, which is initially proposed by Collange, Defour, Graillat, and Iakymchuk in~\cite{Collange15Parco}. ExBLAS is based on combining long accumulators and floating-point expansions in conjuction with error-free transformations. This approach is presented in~\Cref{sec:exblas}. Collange et al. showed~\cite{Collange15Parco} that their algorithms for reproducible and accurate summation have $8$\,\% overhead on $512$ cores (32 nodes) and less than $2$\,\% overhead on 16 cores (one node). While ExSUM covers wide range of architectures as well as distributed-memory clusters, the other routines primarily target GPUs. Exploiting the modular and hierarchical structure of linear algebra algorithms, the ExBLAS approach was applied to construct reproducible LU factorizations with partial pivoting~\cite{Iakymchuk19ReproLU}.

Recently, Mukunoki and Ogita presented their approach to implement reproducible BLAS, called OzBLAS~\cite{ozblas}, with tunable accuracy. This approach is different from both ReproBLAS and ExBLAS as it does not require to implement every BLAS routine from scratch but relies on high-performance (vendor) implementations. Hence, OzBLAS implements the Ozaki scheme~\cite{Ozaki2012} that follows the fork-join approach: the matrix and vector are split (each element is sliced) into sub-matrices and sub-vectors for secure products without overflows; then, the high-performance BLAS is called on each of these splits; finally, the results are merged back using, for instance, the NearSum algorithm. Currently, the OzBLAS library includes dot product, matrix-vector product (gemv), and matrix-matrix multiplication (gemm). These algorithmic variants and their implementations on GPUs and CPUs (only dot) reassure reproducibility of the BLAS kernels as well as make the accuracy tunable up-to correctly rounded results. 

\section{Conclusions and Future Work}
\label{sec:conclusion}
In this work, we addressed the reproducibility of iterative solvers for sparse linear systems using a representative instance of the Preconditioned Conjugate Gradient method. We first analyzed the hybrid MPI + OpenMP tasks implementation of the PCG method and identified two major sources of non-deterministic behavior, namely the \dotp\ product and compiler optimizations. The latter may change the order of operations or replace some of them in favor of the fused multiply-add (\fma) operation. For reproducible and double-layered distributed \dotp\ product, we leveraged the \exblas-approach as well as proposed an alternative lightweight variant based solely on FPEs. Both strategies split the {\tt MPI\_Allreduce} routine into the combination of {\tt MPI\_Reduce} and {\tt MPI\_Bcast}, and perform the intra-node \dotp\ product with FPEs. To tackle compiler interference in computations, we reconstruct computations as well as explicitly invoke \fma\ instructions. Both approaches deliver identical results on two clusters to ensure reproducibility of PCG in the number of iterations, the intermediate and final residuals, the direct errors, as well as the vector-solution on the example of a 3D Poisson's equation with 27 stencil points as well as several test matrices from the  SuiteSparse matrix collection. On a single node, the FPE- and ExBLAS-based reproducible versions of PCG show the maximum overhead of 2.24x and 2.66x, respectively, due to additional memory allocation  and computations. When the communication starts to dominate the execution time, both versions show very low overhead compared with the original non-deterministic implementation: 37.70\,\% for {\em ExBLAS} and 30.50\,\% for {\em Opt} on 768 cores of MareNostrum4; 5.6\,\% for {\em ExBLAS} and 4.6\,\% for {\em Opt} on 128 cores of Tintorrum. This is a solid argument in favor of the reproducible PCG at scale. The code is available at \url{https://github.com/riakymch/ReproCG_MPI_OMP}.

Our study promotes the adoption of reproducibility by design through the proper choice of the underlying libraries as well as a moderate programmability effort. For instance, a brief guidance would be 1) for fundamental numerical computations, to leverage reproducible underlying libraries such as ExBLAS, ReproBLAS, or OzBLAS; and 2) analyze the algorithm and make it reproducible through eliminating any uncertainties that may violate associativity such as reductions and use/ non-use of {\fma}s. Additionally, we argue the need for the bit-wise reproducible and correctly-rounded results for iterative solvers as, nevertheless, they will be enhanced during subsequent iterations as we do not reach the desired tolerance and, thus, do not exploit at full the obtained bit-wise results.

Our future work aims to conduct a deeper analysis of the lightweight approach to support our experimental results. One idea is to bind the length of FPEs to the condition number of the input problem and/or its dynamic range similarly to~\cite{cahi18} for the mixed-precision direct linear solver.

\begin{acks}
To begin with, we would like to thank the reviewers for their valuable comments and suggestions. 
This research was partially supported by the European Union's Horizon 2020 research, innovation programme under the Marie Sk\l{}odowska-Curie grant agreement via the Robust project No. 842528 as well as the Project HPC-EUROPA3 (INFRAIA-2016-1-730897), with the support of the H2020 EC RIA Programme; in particular, the author gratefully acknowledges the support of Vicen\c{c} Beltran and the computer resources and technical support provided by BSC.
The researchers from Universitat Jaume I (UJI) and 
Universitat Polit\'ecnica de Valencia (UPV) were 
supported by MINECO project TIN2017-82972-R. 
Maria Barreda was also supported by the
POSDOC-A/2017/11 project from the Universitat Jaume I.
\end{acks}

\bibliographystyle{SageH}

\end{document}